# Ultra-Strong Gradient Diffusion MRI with Self-Supervised Learning for Prostate Cancer Characterization

Tanishq Patil[1], Snigdha Sen[2], Kieran G. Foley[3], Fabrizio Fasano[4,5], Chantal M. W. Tax[6,7], Derek K. Jones[6], Mara Cercignani[6], Marco Palombo[6,8], Paddy J. Slator[6,8], and Eleftheria Panagiotaki[2]

1. Department of Computer Science, University College London (UCL), London, United Kingdom
2. UCL Hawkes Institute, Department of Computer Science, University College London (UCL), London, United Kingdom
3. Division of Cancer and Genetics, School of Medicine, Cardiff University, Cardiff, United Kingdom
4. Siemens Healthcare Ltd, Camberly, United Kingdom
5. Siemens Healthcare GmbH, Erlangen, Germany
6. Cardiff University Brain Research Imaging Centre (CUBRIC), Cardiff University, Cardiff, United Kingdom
7. Center for Image Sciences, University Medical Center Utrecht, Utrecht, The Netherlands
8. School of Computer Science and Informatics, Cardiff University, Cardiff, United Kingdom

Corresponding Author: Tanishq Patil; tanishq.patil.24@ucl.ac.uk

# Abstract

Diffusion MRI (dMRI) enables non-invasive assessment of prostate microstructure but conventional dMRI metrics such as the Apparent Diffusion Coefficient in multiparametric MRI (mpMRI) are biologically non-specific and reflect a mixture of underlying tissues features rather than distinct histologic characteristics. Integrating dMRI with the compartment-based biophysical VERDICT (Vascular, Extracellular, and Restricted Diffusion for Cytometry in Tumours) framework offers richer microstructural insights, though clinical gradient systems (40–80 mT/m) often suffer from poor signal-to-noise ratio (SNR) at stronger diffusion weightings due to prolonged echo times. Ultra-strong gradients (e.g., 300 mT/m) can mitigate these limitations by improving SNR and contrast-to-noise ratios (CNR). This study investigates whether physics-informed self-supervised VERDICT (ssVERDICT) fitting when combined with ultra-strong gradient data, enhances prostate microstructural characterization relative to current fitting approaches and clinical gradient systems. We developed enhanced ssVERDICT fitting approaches using dense multilayer perceptron (Dense MLP) and convolutional U-Net architectures, benchmarking them against non-linear least-squares (NLLS) VERDICT fitting, original ssVERDICT implementation, and Diffusion Kurtosis Imaging across clinical- to ultra-strong gradient systems. For the same ultra-strong gradient data, Dense ssVERDICT notably outperformed NLLS VERDICT, boosting median CNR by 47%, cutting inter-patient Coefficient of Variation by 52%, and reducing pooled $f_{ic}$ variation by 50%. Overall, Dense ssVERDICT delivered the highest CNR, the most stable parameter estimates, and the clearest tumour-normal contrast compared with conventional fitting methods and clinical gradient systems. These findings underscore that meaningful gains in non-invasive prostate cancer characterization–and consequent reductions in unnecessary biopsies–arise from the combination of advanced gradient systems and deep learning-based modelling. With the recent adoption of ultra-strong gradient MRI in clinical systems, neural network-based parameter estimation can be implemented directly on scanner computers to provide real-time microstructural maps, facilitating seamless translation from research to clinical practice.

# Abbreviations

1. **PCa**: Prostate Cancer
2. **ADC**: Apparent Diffusion Coefficient
3. **DKI**: Diffusion Kurtosis Imaging
4. **VERDICT**: Vascular, Extracellular, and Restricted Diffusion for Cytometry in Tumours
5. **NLLS**: Non-linear Least Squares Fitting
6. **ssVERDICT**: Self-supervised VERDICT
7. **TE**: Echo Time
8. **SNR**: Signal-to-Noise Ratio
9. **CNR**: Contrast-to-Noise Ratio
10. **dMRI**: Diffusion-weighted Magnetic Resonance Imaging
11. **MLP**: Multi-Layer Perceptron
12. **CNN**: Convolutional Neural Network
13. **ROI**: Region of Interest
14. **SP**: Sub-protocol
15. **PGSE**: Pulsed Gradient Spin Echo
16. **EPI**: Echo Planar Imaging
17. **PZ**: Peripheral Zone
18. **PSA**: Prostate-specific Antigen

19. **TRUS**: Transrectal Ultrasound Scan
20. **PT**: Patient
21. **LMFIT**: Non-Linear Least-Squares Minimization and Curve-Fitting
22. **SSR**: Sum of Squared Residuals
23. **MSE**: Mean Squared Error
24. **PReLU**: Parametric Rectified Linear Unit
25. **AICc**: Akaike Information Criterion corrected
26. **BIC**: Bayesian Information Criterion
27. **K**: Excess Kurtosis
28. $\mathbf{f_{ic}}$: Intracellular Volume Fraction
29. **CoV**: Coefficient of Variation
30. **SD**: Standard Deviation
31. **IQR**: Inter-quartile range
32. **PI-RADS**: Prostate Imaging Reporting and Data System
33. **ViT**: Vision Transformers
34. **rVERDICT**: Relaxed VERDICT
35. **VAE**: Variational Autoencoders

# 1 Introduction

Prostate cancer (PCa) is the second most diagnosed cancer in men worldwide [1], commonly assessed through histological examination of biopsied prostate tissue. Although accurate, these procedures are invasive and only sample a tiny fraction of the prostate from within the core of the needles, and therefore, the risk of false negatives is very high [2]. Diffusion MRI offers a non-invasive alternative by probing the diffusion of water molecules to reveal tissue microstructure, and typically covers the whole organ [3]. Historically, diffusion MRI in the prostate has relied on signal representations such as Apparent Diffusion Coefficient (ADC) and Diffusion Kurtosis Imaging (DKI) [4, 5]. While effective for capturing Gaussian and non-Gaussian diffusion respectively, these approaches do not directly relate the MR signal to underlying tumour microstructure.

Biophysical models extend these capabilities by modelling tissue as a set of distinct cellular compartments aiming to link their diffusion properties to key microstructural features like cell density, size, and spatial organization [6]. The Vascular, Extracellular, and Restricted Diffusion for Cytometry in Tumours (VERDICT) model [2] is an advanced biophysical framework that models distinct tissue compartments and can estimate their volume fractions, diffusivities, and cell radius. This framework has shown to produce parameter estimates that are in strong agreement with the ex vivo histological analysis for prostate tissue [7, 8].

Traditionally, parameter estimation for biophysical diffusion models has relied on solving complex inverse problems [9] using non-linear least squares (NLLS) fitting, which is computationally expensive and prone to instability and local minima [10-12]. More recently, supervised deep learning has emerged as a faster and more flexible alternative, although its performance remains constrained by the statistical properties of the training data distribution, leading to biased estimates on unseen test data [13, 14]. Self-supervised learning addresses this limitation by learning directly from the input signal without requiring explicit labels, improving robustness and generalizability [11, 15]. Building on this paradigm, self-supervised VERDICT (ssVERDICT) integrates physics-based [16-18] self-supervised forward models with the VERDICT framework and has demonstrated improved tumour conspicuity and discrimination between normal and cancerous prostate tissue in prior work [11].

Sen et al. (2024) [11] demonstrated the use of ssVERDICT for prostate cancer characterization using 3T clinical diffusion MRI acquired with standard clinical gradient strengths; however, such acquisition systems are restricted in the gradient strength-diffusion time (G–Δ) coverage, and may yield suboptimal image quality and parameter estimation. Recently, Molendowska et al. [19, 20] addressed these limitations by adapting ultra-strong gradients (e.g., 300 mT/m), which had previously been restricted to head only systems, for prostate imaging. This enabled dMRI acquisition at higher b-values with significantly shorter echo times (TE), and improved signal-to-noise ratio (SNR) and contrast-to-noise ratio (CNR), particularly for early-stage prostate cancer. Their study established that high-gradient diffusion MRI is both technically feasible and clinically beneficial for prostate cancer.

This study aims to apply physics-informed self-supervised VERDICT fitting to the same ultra-strong gradient dMRI, leveraging the synergy between advanced acquisition and biophysical modelling to enhance non-invasive microstructural characterization of prostate cancer. The adoption of ultra-strong gradient acquisition systems has until recently been largely confined to research settings, but with their emerging availability in commercial clinical scanners, it is timely to evaluate how the VERDICT framework performs under these acquisitions and whether such settings offer any additional sensitivity for prostate microstructural characterization compared with conventional clinical gradient strengths. We hypothesize that ssVERDICT with ultra-strong gradients will yield higher lesion conspicuity and more stable parameter estimates than NLLS VERDICT and DKI at clinical gradient strengths.

This work represents the first investigation of integrating ultra-strong gradient dMRI with VERDICT modelling in the context of prostate cancer, thereby providing a novel contribution to the field. We designed deep learning models, including Multi-Layer Perceptron-based (MLP) and Convolutional Neural Networks-based (CNN) architectures, and performed comprehensive hyperparameter optimization to obtain the optimal configuration for ssVERDICT model fitting. We show that the self-supervised VERDICT model with a dense MLP architecture produces low-error, noise-free reconstructions of diffusion MRI signals, yielding accurate and stable parameter estimates that enhance tumour-normal tissue discrimination within regions of interest (ROI) compared to conventional clinical methods.

# 2 Materials and Methods

## 2.1 Data

### 2.1.1 Data Acquisition

MRI data were acquired on a Connectom research-only 3T MAGNETOM Skyra scanner (Siemens Healthcare, Erlangen, Germany) equipped with a 300 mT/m gradient coil [19]. A pulsed gradient spin echo (PGSE) with echo-planar imaging (EPI) sequence was used to acquire the data. From this single acquisition, three dMRI sub-protocols (SP) were derived, where each used the same sampling scheme (b-values and encoding directions; see below), but differed with maximum allowed gradient amplitude, diffusion waveforms timings and echo time (TE). The gradient amplitudes reflected a high-performance system with 300 mT/m (sub-protocol 1, SP1), and clinical systems with 80 mT/m (SP2) and 40 mT/m (SP3), respectively. Each was acquired with corresponding minimum achievable TE for the maximum b-value, resulting in TE/$\delta$/$\Delta$ [ms]: SP1 – 54/5/25, SP2 – 70/16/32, SP3 – 95/26/48, respectively, with TE, $\delta$, and $\Delta$ held constant across all b-shells within each sub-protocol. The following imaging parameters were the same across sub-protocols: b = 0.05, 0.5, 1.5, 2.0, 3.0 ms/μm$^2$, each sampled at 15 non-collinear directions, with interspersed non-diffusion-weighted (b = 0 ms/μm$^2$) volumes, in-plane resolution = 1.3×1.3 mm$^2$, FOV = 220×220 mm$^2$, matrix = 168×168, slice thickness = 5 mm, 14 slices, GRAPPA = 2, SMS = 2, partial Fourier = 6/8, bandwidth = 1860 Hz/pixel. Total scan time was approximately 20 min (6 min 35 sec per sub-protocol).

### 2.1.2 Patient Cohort

The dataset represents a subset of participants examined in [19], with the ethical approval for the study granted by the Cardiff University School of Psychology REC and the NHS REC (ref: 20/OCT/8264). This study included four healthy male volunteers (46–64 years; mean age = 54.8 ± 7.9 years) and five patients with histologically confirmed prostate adenocarcinoma enrolled in the Active Surveillance Programme (67–74 years; mean age = 69.8 ± 2.7 years). Cancerous lesions in the peripheral zone (PZ) were verified through elevated PSA levels, multiparametric MRI, and TRUS-guided biopsy, with Gleason scores of 3+3 or 3+4. Two of the five patients (hereafter labelled as PT 1, and PT 2) were assigned a Gleason score of 3+3, and the remaining three patients (hereafter labelled as PT 3 – PT 5) were assigned a score of 3+4. All participants underwent MRI safety checks and gave written informed consent.

### 2.1.3 Data Preprocessing

Prostate masks were manually generated by delineating the gland and corresponding tumour ROIs on each patient's images, using radiologist-confirmed tumour locations [19] as reference. The pre-processing pipeline in [19] involved denoising [21], Gibbs ringing correction [22], partial distortion correction [23, 24], and adjustment of effective b-values and gradient directions [25], without eddy-current or motion correction. Subsequently, further preprocessing was performed on the dMRI data to enable training and inference with deep learning models. The voxel-wise b-values were computed as the Euclidean-norm of the unnormalized gradient direction vectors. Due to voxel-wise gradient non-uniformity in the dataset, effective b-values varied spatially across the imaging volume. To maintain consistent input dimensions for deep learning-based training, voxel-wise b-values were approximated to their closest theoretical values (b = 0–3 ms/µm²), and the volumes with same theoretical b-values were spatially averaged. For NLLS and MLP models, prostate masks were applied, spatial dimensions flattened, voxels with zero signal values discarded, and all signals normalized with averaged b = 0 ms/µm² volume, producing 2D arrays of [number of voxels in a volume × theoretical b-values] concatenated across subjects. For CNNs, masks were omitted to preserve spatial uniformity. The images were kept at original resolution, spatially averaged across identical b-value volumes, normalized, and permuted to [slices × b-values × image height × image width].

### 2.1.4 Data Split

Each subject underwent a single dMRI acquisition from which three sub-protocols were derived. An identical pre-processing and data-splitting approach was applied across sub-protocols to maintain consistency in analysis. To emulate a clinical workflow, subjects were partitioned into three sets in chronological order of their acquisition date:

- Training Set: 2 healthy controls, 3 patients: PT 1 – PT 3
- Validation Set: 1 healthy control, 1 patient: PT 5
- Test Set: 1 healthy control, 1 patient: PT 4

These splits enabled assessment of training quality and evaluation on unseen data, allowing us to determine whether the deep learning models learned meaningful mappings for estimating the parameters underlying the measured diffusion signals. Although the model could in principle be trained and evaluated on individual subjects, as is done with NLLS, this would obscure potential memorization or overfitting and hinder assessment of whether the model truly learned the inverse mapping of the signal representations.

## 2.2 Diffusion MRI Frameworks

### 2.2.1 Diffusion Kurtosis Imaging

The powder-averaged DKI formulation extends the mono-exponential ADC with an additional second-order term accounting for non-Gaussian diffusion due to tissue heterogeneity [5] and complexity. The normalized diffusion signal is defined as [26]:

$$\frac{S(b)}{S_0} = \exp\left(-bD_k + \frac{1}{6}b^2 D_k^2 K\right) \tag{2.1}$$

where $b$ is the diffusion weighting (b-value), $S(b)$ and $S_0$ are signals at diffusion weighting b and zero respectively, $D_k$ is the diffusivity and $K$ is the excess kurtosis. Fitting the DKI representation on the dMRI data results in the estimation of two parameters – $D_k$ and $K$. All b-values up to b = 3.0 ms/µm² were included after confirming that the corresponding normalised signal remained within the commonly accepted convergence radius of the DKI representation for prostate tissue.

### 2.2.2 VERDICT

The VERDICT framework is formulated as the sum of diffusion signal contributions from intracellular (IC), extracellular-extravascular (EES), and vascular (VASC) compartments. The total normalized diffusion-weighted signal is given by [2, 11, 27]:

$$\frac{S}{S_0} = f_{ic}S_{ic}(d_{ic}, R, b, \Delta, \delta) + f_{ees}S_{ees}(d_{ees}, b) + f_{vasc}S_{vasc}(d_{vasc}, b) \quad (2.2)$$

where $f_i, S_i$, and $d_i$ are the volume fractions, normalized diffusion signals, and diffusivities respectively in biophysical compartment $i$ ($i = ic, ees$, or $vasc$); $b$ is the diffusion weighting (b-value), $S_0$ is the signal at $b = 0$, $R$ is the mean cell radius, $\Delta$ is the separation of the gradient waveform, and $\delta$ is the duration of the gradient waveform. Furthermore, the conditions $\sum_{i=1}^{3} f_i = 1$ and $0 \leq f_i \leq 1$ allow us to compute $f_{vasc}$ directly from the values of $f_{ic}$ and $f_{ees}$, given $f_{vasc} + f_{ic} + f_{ees} = 1$. In accordance with earlier studies [26], $d_{ic}$ was set to 2 µm$^2$/ms, while $d_{vasc}$ was set to 8 µm$^2$/ms. Therefore, the VERDICT framework requires estimation of four key microstructural parameters during model fitting [11, 27] – $f_{ic}$, $f_{ees}$, $R$, and $d_{ees}$.

## 2.3 Fitting Strategies

### 2.3.1 Non-linear least-squares

NLLS parameter estimation was performed using the LMFIT library (v1.3.3) in Python, which applies the Levenberg-Marquardt optimization algorithm, as in [2, 4, 26]. The objective function was defined as the difference between the measured diffusion signal and the model-predicted signal, and the sum of squared residuals (SSR) was minimized to obtain the model fitting. This yielded voxel-wise SSR values corresponding to the best-fit parameter estimates.

### 2.3.2 Deep Learning MLP: Baseline Architecture

We use the previously published ssVERDICT architecture as our starting point. ssVERDICT comprises a self-supervised autoencoder with a physics-informed decoder that minimizes the mean-squared error (MSE) between the normalized noisy measured MR signals ($S_{ip}$: input data) and the normalized noise-free signals reconstructed from the estimated latent space parameters ($\hat{S}_{pred}$: output data). This implementation, hereafter referred to as Baseline MLP, follows the architecture introduced in [11].

For both DKI and VERDICT, we implemented this architecture with an MLP encoder that estimated the latent parameters and a decoder, explicitly defined by corresponding differentiable model equations (eq. 2.1 and 2.2). The input layer of the encoder was set to size 6, with each node corresponding to a distinct acquisition b-value – (0, 0.05, 0.5, 1.5, 2, 3) ms/µm$^2$. The remainder of the architecture followed the previously published design: three hidden layers with 10 neurons each and a latent output layer of size 2 for DKI, and 4 for VERDICT. All hidden layers employed Parametric Rectified Linear Unit (PReLU) activations. The dropout layer ($p = 0.2$) was relocated from the input layer in original ssVERDICT to immediately before the output layer to improve the stochastic regularization and mitigate overfitting.

To ensure non-negativity while preserving smooth gradient propagation, encoder outputs were passed through a Softplus activation and then clamped to physiologically plausible parameter-specific bounds. The constrained parameters were then supplied to the decoder, to reconstruct biophysically meaningful, normalized, noise-free diffusion signals. The MSE between reconstructed and measured signals [12] served as the loss function. Training minimized this loss using the Adam optimizer, with gradients backpropagated end-to-end from the decoder to the encoder. The network employed the same hyperparameter configuration reported in [11], with a batch size of 256 and a constant learning rate of 0.001. A fixed number of epochs was used to systematically monitor training and validation loss and assess model convergence. Preliminary monitoring indicated that the VERDICT Baseline MLP required approximately 300 epochs for the loss to stabilise, whereas the less complex DKI representation converged within about 60 epochs. These hyperparameters were applied uniformly across all three gradient strength datasets.

### 2.3.3 Deep Learning MLP: Dense and Improved Architecture

The Dense MLP extended the baseline architecture by deepening the encoder to maximize the model's ability to capture information from all considered diffusion-weighted data volumes. The latent space and the decoder remained unchanged. The encoder consisted of an input layer of 6 neurons followed by five hidden layers with 32, 64, 128, 64, and 32 neurons, respectively. This deeper design increased the model's representational capacity, raising the number of trainable VERDICT parameters from 337 (315 for DKI) in the baseline to 21,129 (21,063 for DKI) in the dense variant. PReLU activations and dropout layer were kept the same as in the baseline architecture. Unlike the baseline, the dense model applied a Sigmoid activation at the encoder output to constrain parameters to [0, 1], which were then linearly scaled to their biophysically realistic ranges. This approach preserved smooth gradient flow, in contrast to clamping, which can hinder gradients near the bounds. A schematic of the VERDICT dense MLP architecture is shown in Figure 1.

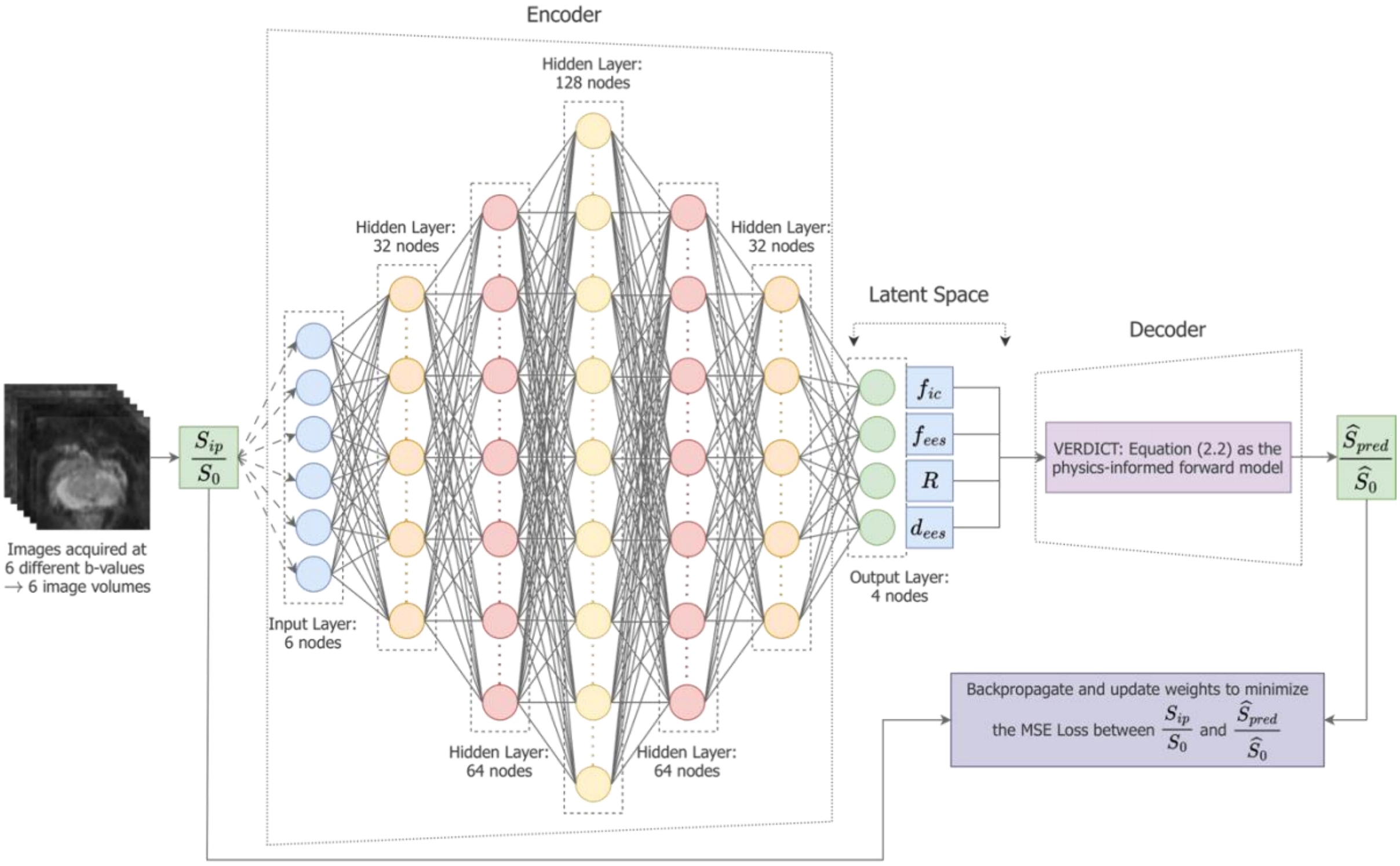

**Figure 1:** Schematic of VERDICT Dense MLP Architecture: The network takes six normalized diffusion-weighted signals as input, maps them to a four-node latent space representing VERDICT parameters, and passes these through a physics-

informed decoder based on the VERDICT equation (eq. 2.2) to reconstruct the diffusion-weighted signal and minimize reconstruction error.

A full factorial grid search was performed over batch size (DKI ∈ [32, 64, 128, 256]; VERDICT ∈ [64, 128, 256, 512]) and initial learning rate ([1e-4, 1e-3, 1e-2, 1e-1], common to both), with tuning repeated independently for all three gradient strength datasets. The number of epochs was fixed at 60, as both VERDICT and DKI implemented with the Dense MLP architecture converged efficiently within this range, in contrast to the VERDICT baseline implementation which required approximately 300 epochs. A learning rate scheduler reduced the initial learning rate by a factor of 0.1 every 10 epochs, enabling more precise weight updates and improved loss minimization. Table 1 lists the optimal hyperparameters identified for the DKI and VERDICT Dense MLP architectures.

### 2.3.4 Deep Learning CNN: U-Net Architecture

This procedure was not applied to DKI due to training instability with the available diffusion MRI data.

Within the similar physics-informed encoder-decoder framework, U-Net architecture was used as an encoder with 6 input channels, each channel corresponding to a 2D image slice. The U-Net comprised a symmetric encoder-decoder structure with convolutional blocks and skip connections, compressing the input diffusion MRI slices to a 10×10 latent representation and progressively reconstructing them to the original 168×168 resolution [10, 15, 28]. The final 1×1 convolution mapped the 64-channel feature representation to the four VERDICT parameter maps, enabling spatially informed parameter estimation. This final convolution used a Sigmoid activation to constrain outputs to [0, 1], which were rescaled to physical ranges, flattened, and passed to the physics-informed VERDICT decoder to reconstruct the diffusion-weighted signals, with MSE minimized via Adam optimizer.

Unlike MLPs, the U-Net implementation omitted prostate masks during preprocessing to maintain uniform image size, while applying the mask only during loss calculation so that gradient updates were restricted to prostate voxels. This avoids zero signal background bias, reduces zero-activation issues in PReLU layers, and improves both computational efficiency and convergence.

To ensure stable training with hyperparameter tuning, the U-Net batch size range was reduced, as channel-wise image processing in CNNs yields fewer effective training samples. A comprehensive grid search was performed over batch sizes of [1, 2, 5, 10], and initial learning rate of [1e-4, 1e-3, 1e-2, 1e-1]. The model was trained for 60 epochs at each iteration with a learning rate scheduler decreasing the initial value by a factor of 0.5 every 5 epochs. Table 1 lists the optimal hyperparameter obtained for VERDICT U-Net.

**Table 1:** Optimal hyperparameter configurations for Dense MLP (DKI and VERDICT) and U-Net (VERDICT) architectures across the three sub-protocols.

| Diffusion MRI Frameworks | Sub-Protocol (Gradient Strength) | Optimal Batch Size | Optimal Learning Rate (initial) |
|---|---|---|---|
| DKI: Dense MLP | SP1 (300 mT/m) | 64 | 0.01 |
| | SP2 (80 mT/m) | 64 | 0.01 |
| | SP3 (40 mT/m) | 128 | 0.01 |
| VERDICT: Dense MLP | SP1 (300 mT/m) | 128 | 0.01 |
| | SP2 (80 mT/m) | 256 | 0.01 |
| | SP3 (40 mT/m) | 128 | 0.01 |
| VERDICT: U-Net | SP1 (300 mT/m) | 5 | 0.001 |
| | SP2 (80 mT/m) | 5 | 0.001 |
| | SP3 (40 mT/m) | 5 | 0.001 |

### 2.3.5 Evaluation Strategies and Metrics

To evaluate the quality of fit to the data with different frameworks and fitting strategies, we computed the MSE between the normalized input and the noise-free reconstructed diffusion signals. The Akaike Information Criterion corrected (AICc) and Bayesian Information Criterion (BIC) were used to compare the models and assess whether reductions in loss in more complex models were achieved without overparameterization for the ultra-strong gradient data acquired under sub-protocol SP1. To differentiate malignant from normal tissues, biomarker ($K$ & $f_{ic}$) differences between paired tumour and normal ROIs (drawn from the same prostate) were tested for normality using the Shapiro-Wilk test, followed by a paired t-test or, if normality was violated, a Wilcoxon's signed rank test.

We computed the mean biomarker value within the tumour ROI for each patient, and derived the Coefficient of Variation (CoV) as the ratio of standard deviation to the mean of these patient-level averages to assess inter-patient variability and model robustness. We additionally reported the pooled voxel-wise standard deviation (SD) of biomarker values across all patients to characterize voxel-level heterogeneity and instability (salt-and-pepper noise). Finally, we computed the CNR (median and inter-quartile range [IQR]) for each fitted model across the patients to quantify lesion conspicuity relative to contralateral normal tissue, following the approach in [7, 29].

For Gleason grade discrimination, the independent samples (drawn from different prostates) were first tested for normality using Shapiro-Wilk test, and when violated, assessed for variance homogeneity (Levene's and Bartlett's tests). Accordingly, biomarker differences were then evaluated using the non-parametric Mann-Whitney U test.

# 3 Results

## 3.1 Model Evaluation

Figures 2 and 3 show the training and validation loss curves corresponding to optimal hyperparameter configurations for each model training process. All curves show clear evidence of stable convergence with VERDICT deep learning models saturating at lower MSE values than their DKI counterparts, with U-Net as an exception. Additionally, sub-protocol SP1 exhibits lower saturation MSE values than SP2 and SP3 for each fitting method.

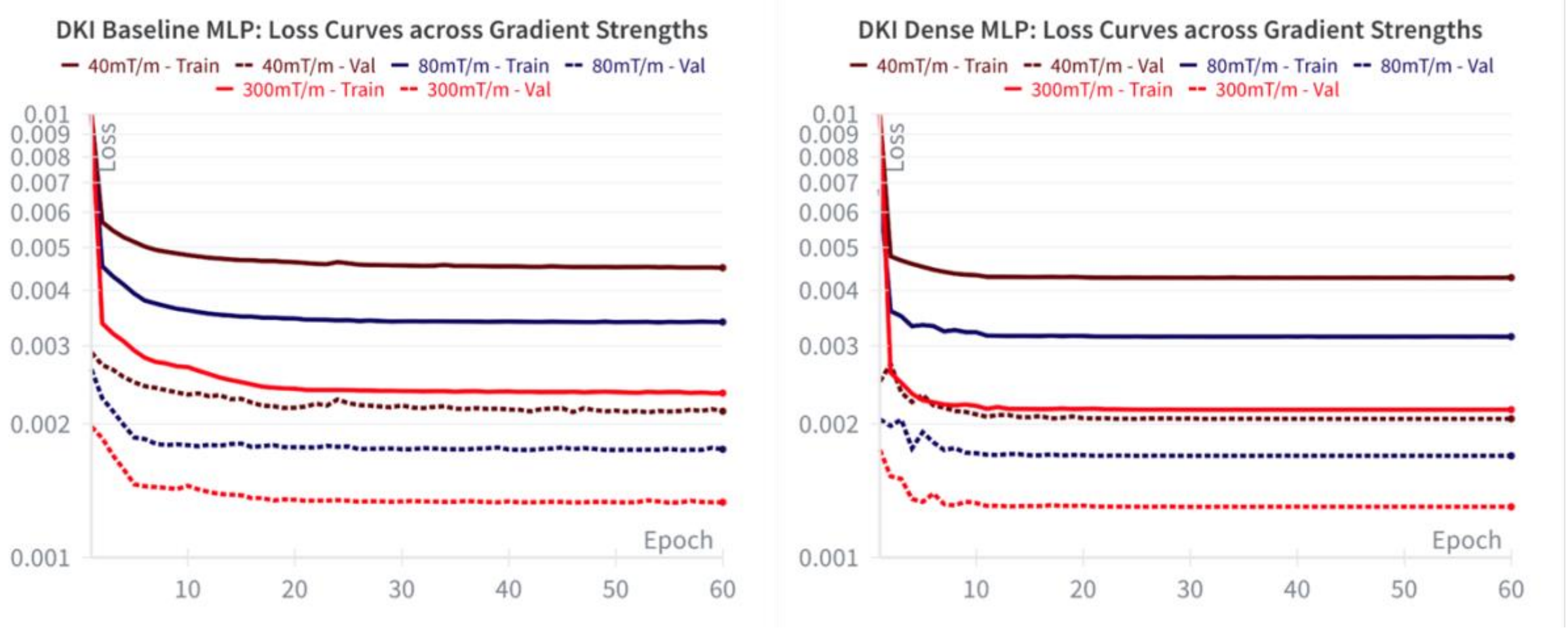

**Figure 2:** DKI Loss Curves: Training and Validation loss curves for DKI Baseline MLP and Dense MLP models across three gradient sub-protocols. Both models show stable convergence, with Dense MLP consistently saturating at lower loss values

than Baseline MLP. Loss decreases with increasing gradient strength, with SP1 (300 mT/m) yielding the lowest final MSE, followed by SP2 (80 mT/m) and SP3 (40 mT/m). The y-axis (Loss) is displayed on a logarithmic scale.

Tables 2–4 provide a comparison of MSE values for patients and healthy controls on the held-out test set along with the validation and training losses at the best-performing epoch across SP1, SP2 and SP3 respectively. These results were obtained using model weights from the epoch with the lowest validation loss on the held-out test set.

Across all three sub-protocols, the VERDICT Dense MLP consistently achieved the lowest MSE among the deep learning models (Baseline MLP, Dense MLP, and U-Net) on the training and validation sets. Within the unseen test data, the best performing method varied by subject with the VERDICT Dense MLP producing the lowest errors for the Healthy Control, and VERDICT NLLS achieving the lowest MSE for PT 4. However, NLLS required substantially longer computation time (808 s for PT 4; 1709 s for Healthy Control) compared to neural network inference (2 s for PT 4; 2.4 s for Healthy Control), with a training cost of 396 s across 5 subjects. These models marginally outperformed the next-best approaches across their respective cases in the test set.

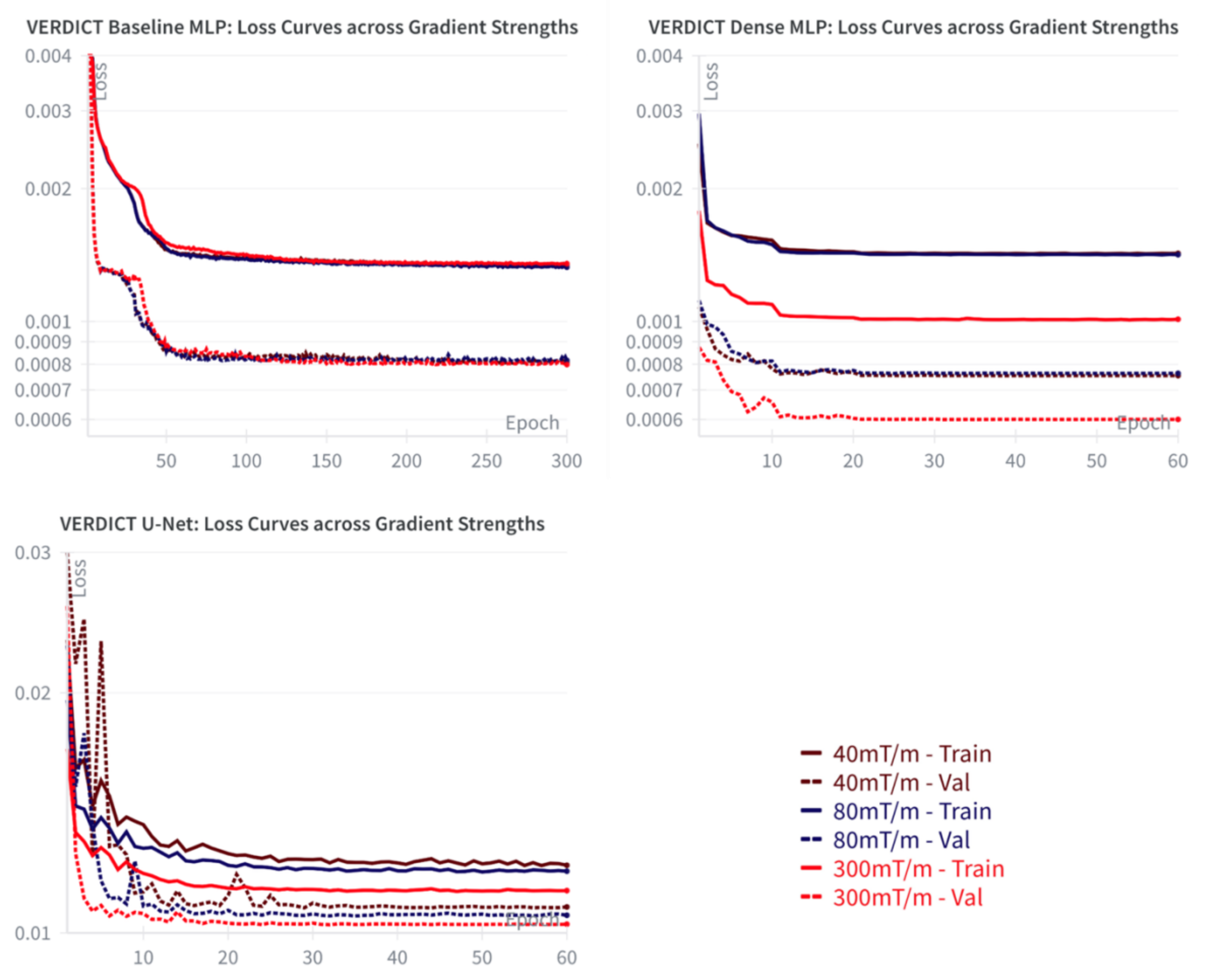

**Figure 3:** VERDICT Loss Curves: Training and Validation loss curves for VERDICT Baseline MLP, Dense MLP, and U-Net models across three gradient sub-protocols. The Baseline MLP requires 300 epochs to converge, with losses nearly identical across SP1–SP3, reflecting limited representational capacity. Dense MLP and U-Net converge within 60 epochs. Dense MLP achieved lower loss, particularly for SP1, demonstrating improved learning and effective use of high-gradient data, while SP2 and SP3 remain slightly lower but still outperform Baseline MLP. Notably, U-Net exhibits higher losses relative to Baseline MLP and Dense MLP, though it still perform best on the ultra-strong gradient SP1 dataset relative to SP2 and SP3. The y-axis (Loss) is displayed on a logarithmic scale.

**Table 2:** SP1 (300 mT/m): Training, validation, and test losses across diffusion MRI frameworks and fitting methods. On both the training and validation sets, VERDICT Dense MLP achieves the lowest reconstruction losses among the deep learning models. In the test set, VERDICT NLLS for Patient 4 and the VERDICT Dense MLP for the Healthy Control yield the lowest losses, marginally outperforming the next-best models.

| **Diffusion MRI Frameworks** | **Fitting Models** | **Model Training** | | | **Model Inference** | |
|---|---|---|---|---|---|---|
| | | **Best Val Epoch** | **Val Loss** | **Train Loss** | **Test Loss: Patient 4** | **Test Loss: Healthy** |
| DKI | NLLS | — | — | — | 6.71e-4 | 2.18e-3 |
| | Baseline MLP | 55 | 1.33e-3 | 2.36e-3 | 6.96e-4 | 2.25e-3 |
| | Dense MLP | 30 | 1.30e-3 | 2.15e-3 | 6.73e-4 | 2.21e-3 |
| VERDICT | NLLS | — | — | — | **2.62e-4** | 9.99e-4 |
| | Baseline MLP | 261# | 7.99e-4 | 1.35e-3 | 4.07e-4 | 1.12e-3 |
| | Dense MLP | 39 | **6.01e-4** | **1.01e-3** | 2.74e-4 | **9.73e-4** |
| | U-Net | 46 | 1.02e-2 | 1.13e-2 | 7.41e-3 | 3.02e-2 |

**Table 3:** SP2 (80 mT/m): Training, validation, and test losses across diffusion MRI frameworks and fitting methods. On both the training and validation sets, VERDICT Dense MLP achieves the lowest reconstruction losses among the deep learning models. In the test set, VERDICT NLLS for Patient 4 and the VERDICT Dense MLP for the Healthy Control yield the lowest losses, marginally outperforming the next-best models.

| **Diffusion MRI Frameworks** | **Fitting Models** | **Model Training** | | | **Model Inference** | |
|---|---|---|---|---|---|---|
| | | **Best Val Epoch** | **Val Loss** | **Train Loss** | **Test Loss: Patient 4** | **Test Loss: Healthy** |
| DKI | NLLS | — | — | — | 1.04e-3 | 2.97e-3 |
| | Baseline MLP | 56 | 1.74e-3 | 3.39e-3 | 1.11e-3 | 3.03e-3 |
| | Dense MLP | 30 | 1.69e-3 | 3.15e-3 | 1.04e-3 | 3.00e-3 |
| VERDICT | NLLS | — | — | — | **4.31e-4** | 1.24e-3 |
| | Baseline MLP | 205# | 9.65e-4 | 1.79e-3 | 5.97e-4 | 1.43e-3 |
| | Dense MLP | 28 | **7.63e-4** | **1.42e-3** | 4.49e-4 | **1.23e-3** |
| | U-Net | 46 | 1.05e-2 | 1.21e-2 | 7.84e-3 | 3.06e-2 |

**Table 4:** SP3 (40 mT/m): Training, validation, and test losses across diffusion MRI frameworks and fitting methods. On both the training and validation sets, VERDICT Dense MLP achieves the lowest reconstruction losses among the deep learning models. In the test set, VERDICT NLLS for Patient 4 and the VERDICT Dense MLP for the Healthy Control yield the lowest losses, marginally outperforming the next-best models.

| **Diffusion MRI Frameworks** | **Fitting Models** | **Model Training** | | | **Model Inference** | |
|---|---|---|---|---|---|---|
| | | **Best Val Epoch** | **Val Loss** | **Train Loss** | **Test Loss: Patient 4** | **Test Loss: Healthy** |
| DKI | NLLS | — | — | — | 3.18e-3 | 4.79e-3 |
| | Baseline MLP | 53 | 2.12e-3 | 4.52e-3 | 3.20e-3 | 4.87e-3 |
| | Dense MLP | 25 | 2.05e-3 | 4.28e-3 | 3.18e-3 | 4.82e-3 |
| VERDICT | NLLS | — | — | — | **2.05e-3** | 1.92e-3 |
| | Baseline MLP | 291# | 9.96e-4 | 1.91e-3 | 2.31e-3 | 2.13e-3 |
| | Dense MLP | 36 | **7.54e-4** | **1.44e-3** | 2.06e-3 | **1.92e-3** |
| | U-Net | 43 | 1.08e-2 | 1.23e-2 | 1.02e-2 | 3.16e-2 |

#: VERDICT Baseline MLP ran for 300 epochs unlike 60 for the other methods.

—: Fields that are not applicable for the NLLS method.

Table 5 summarises the AICc and BIC values and corresponding ranks for all diffusion MRI frameworks and fitting strategies evaluated on the patient data in the test set (PT 4). Both criteria yielded an identical ordering of model performance, with VERDICT NLLS emerging as the best-fitting approach and the VERDICT Dense MLP ranking a close second. Across all comparisons, the VERDICT-based models performed better than the corresponding DKI-based models, indicating that the additional microstructural detail captured by VERDICT provides meaningful information rather than simply increasing the model complexity.

**Table 5:** SP1 (300 mT/m): AICc and BIC scores and ranks for different combinations of diffusion MRI frameworks and fitting methods, evaluated on the Patient 4 from the test split of the ultra-strong gradient data. VERDICT NLLS marginally outperforms the VERDICT Dense MLP, achieving the best AICc and BIC ranks. Notably, the VERDICT-based models consistently outperform their DKI counterparts.

| **Models** | **AICc ($* 10^3$)** | **Rank** | **BIC ($* 10^3$)** | **Rank** | **No. of Parameters** |
|---|---|---|---|---|---|
| VERDICT: NLLS | -68.542 | 1 | -68.514 | 1 | 4 |
| VERDICT: Dense MLP | -68.170 | 2 | -68.142 | 2 | 4 |
| VERDICT: Baseline MLP | -64.881 | 3 | -64.853 | 3 | 4 |
| DKI: NLLS | -60.729 | 4 | -60.715 | 4 | 2 |
| DKI: Dense MLP | -60.704 | 5 | -60.690 | 5 | 2 |
| DKI: Baseline MLP | -60.425 | 6 | -60.411 | 6 | 2 |
| VERDICT: U-Net | -40.761 | 7 | -40.733 | 7 | 4 |

## 3.2 Parameter Maps and Tumour-Normal Tissue Discrimination

All deep learning models were successfully trained with no evidence of overfitting, as demonstrated by the stable validation loss curves shown in Figures 2 and 3. This indicates that the models effectively approximated the inverse formulations of DKI and VERDICT, enabling inference on training and validation subject data to generate estimated parameter maps. In-vivo DKI and VERDICT parameter maps were generated for the five patients to qualitatively assess fitting quality and evaluate tumour ROI conspicuity, with two cases discussed below. Among these, the VERDICT intracellular volume fraction ($f_{ic}$) and DKI excess kurtosis ($K$) were obtained, both established as effective non-invasive biomarkers for cancer diagnosis [26, 30, 31].

Figure 4 illustrates the in vivo maps of fitted $K$ and $f_{ic}$ biomarkers for Patient 5. DKI maps showed little variability in visual quality across fitting methods, reflecting the low representational capacity of DKI and its ability to capture only limited tumour microstructural information. As a result, all fitting approaches converged to similar solutions regardless of complexity, unlike in VERDICT. VERDICT with Dense MLP and U-Net at 300 mT/m (SP1) exhibited superior lesion discrimination and fitting quality compared with results at sub-protocols SP2 and SP3, with maps from SP3 showing the poorest performance. VERDICT NLLS maps remained uninformative, with multiple spurious highlighted regions reflecting the method’s susceptibility to noise.

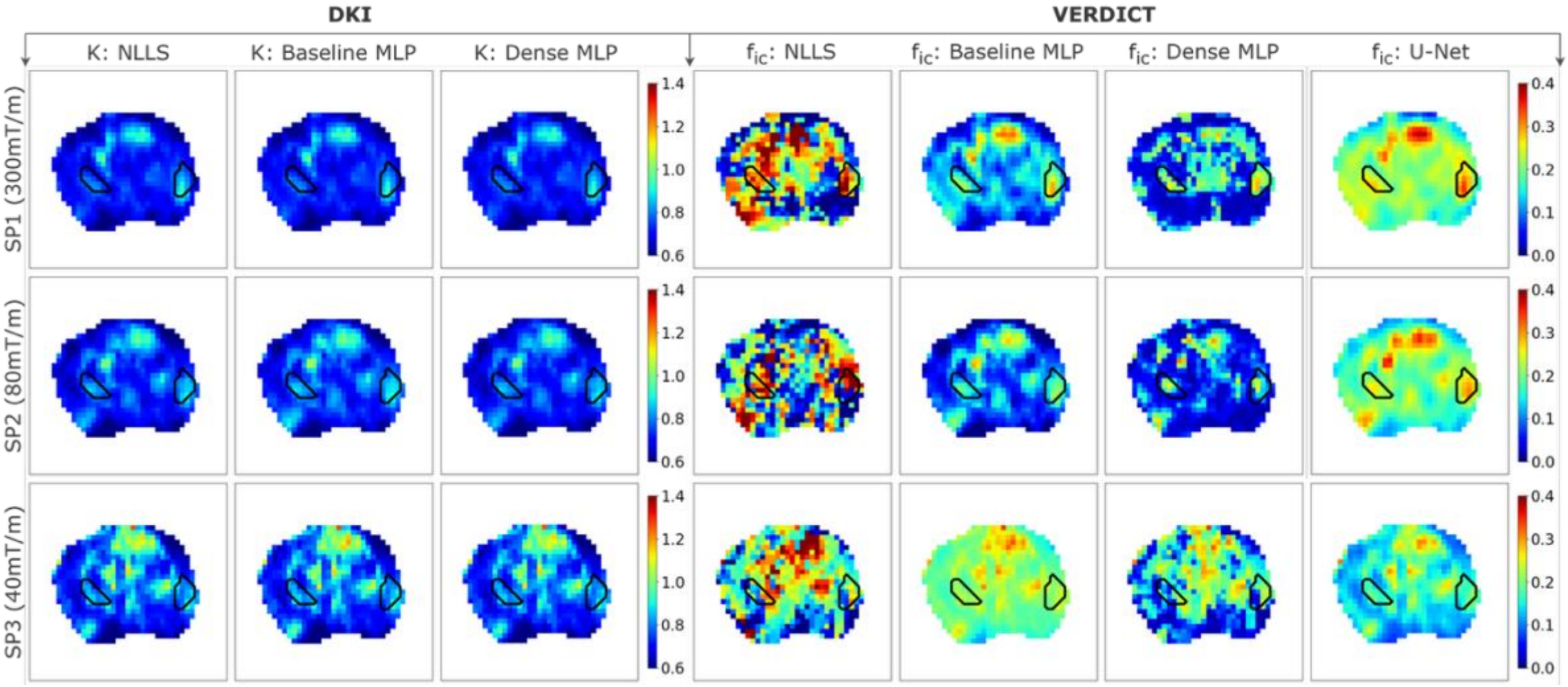

**Figure 4:** Comparison of $K$ and $f_{ic}$ parameter maps obtained using different fitting methods for both DKI and VERDICT fitting models across the three sub-protocols for Patient 5. There are two 3+4 Gleason grade tumours present, one at left lateral PZ, and the other at right lateral PZ, both at mid-gland level. Deep learning methods (Baseline MLP, Dense MLP, U-Net) noticeably improve fitting over NLLS, with the best lesion conspicuity observed under SP1 at 300 mT/m. Dense MLP and U-Net slightly outperform Baseline MLP, while results under SP2 and SP3 remain suboptimal, underscoring the value of ultra-strong gradients for resolving tumour microstructure. For consistent visualization, $K$ values are constrained to the interval [0.6, 1.4], and $f_{ic}$ values are restricted to [0, 0.4], as indicated by the corresponding colorbars.

Figure 5 shows boxplots of the estimated $K$ and $f_{ic}$ distributions for Patient 5. All fitting methods distinguished malignant from normal tissue with strong statistical significance ($p < 0.001$) under SP1. For SP2, malignant–normal separation was still observed, but some methods showed weaker significance ($p < 0.05$). The results with data acquired under SP3 showed poorest performance, with median $K$ and $f_{ic}$ values in some normal tissues exceeding those of lesions, which is in contrast to established literature [7, 27]. Despite the lesion's small size, the parameter values for SP1 reliably distinguished tumour from normal tissue and exhibited narrower biomarker ranges.

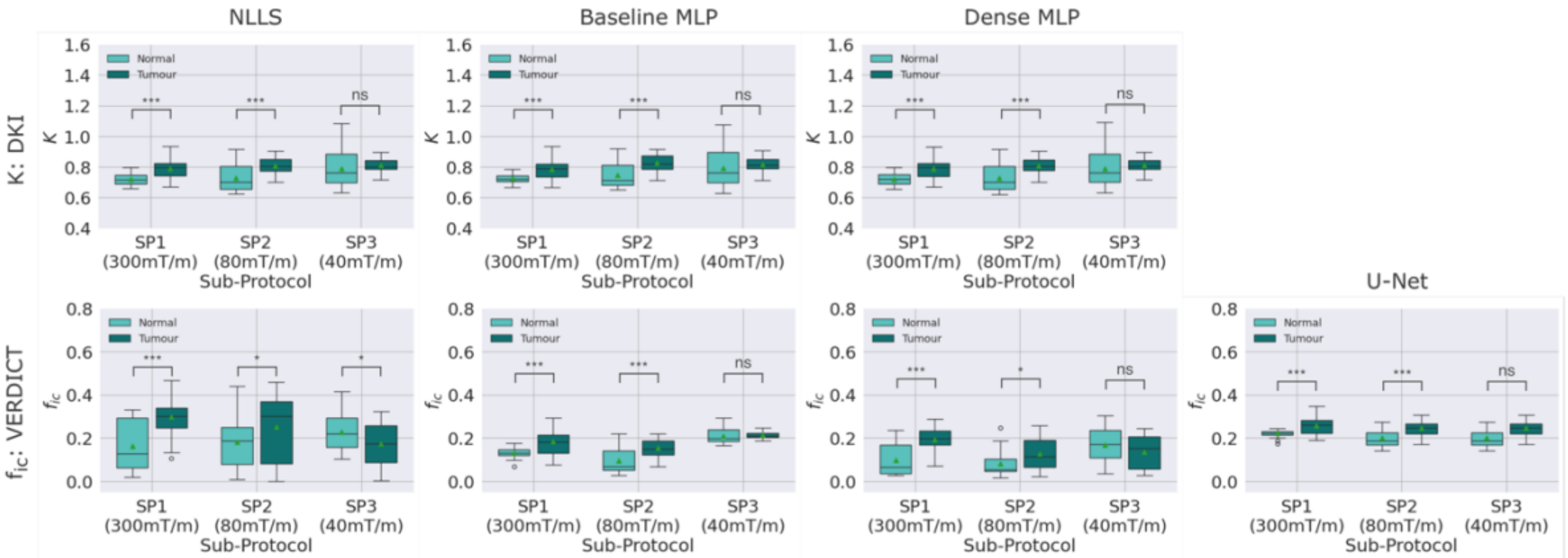

**Figure 5:** Boxplots illustrating the distributions of $K$ and $f_{ic}$ values across the three sub-protocols, compared across different fitting methods for both DKI and VERDICT fitting models for Patient 5. In DKI, data acquired under SP1 and SP2 show significant separation between tumour and normal voxels, whereas data under SP3 do not, yielding implausible normal tissue $K$ values that exceed those in tumour tissue. For VERDICT, strong tumour-normal separation ($p < 0.001$) is observed under SP1 across all methods, while under SP2, the separation is weaker but remains significant. Results obtained under SP3 exhibit poor performance with physiologically unrealistic results for $f_{ic}$ , similar to behaviour observed for DKI. To ensure consistency in visualization, the $K$ values are constrained within the interval [0.4, 1.6], while the $f_{ic}$ values are restricted to [0, 0.8]. The following notation indicates statistical significance: $*** \rightarrow p < 0.001$; $** \rightarrow p < 0.01$; $* \rightarrow p < 0.05$; and *ns (non-significant)* $\rightarrow p > 0.05$.

Figure 6 shows the estimated $K$ and $f_{ic}$ parameter maps for Patient 3. Across all DKI fittings, both lesions were most conspicuous under sub-protocol SP1, while SP2 and SP3 were associated with reduced visibility and more ambiguous boundaries. VERDICT Baseline MLP, Dense MLP, and U-Net were able to effectively capture the lesion conspicuity with Dense MLP providing the best visual contrast. VERDICT NLLS maps were noisy and showed insufficient contrast to distinguish lesion boundaries from surrounding tissue. Parametric maps for SP3 showed minimal discrimination and displayed artefacts in the upper right regions of the maps, pertaining to elevated levels of $K$ and $f_{ic}$ values that were unrelated to tumour.

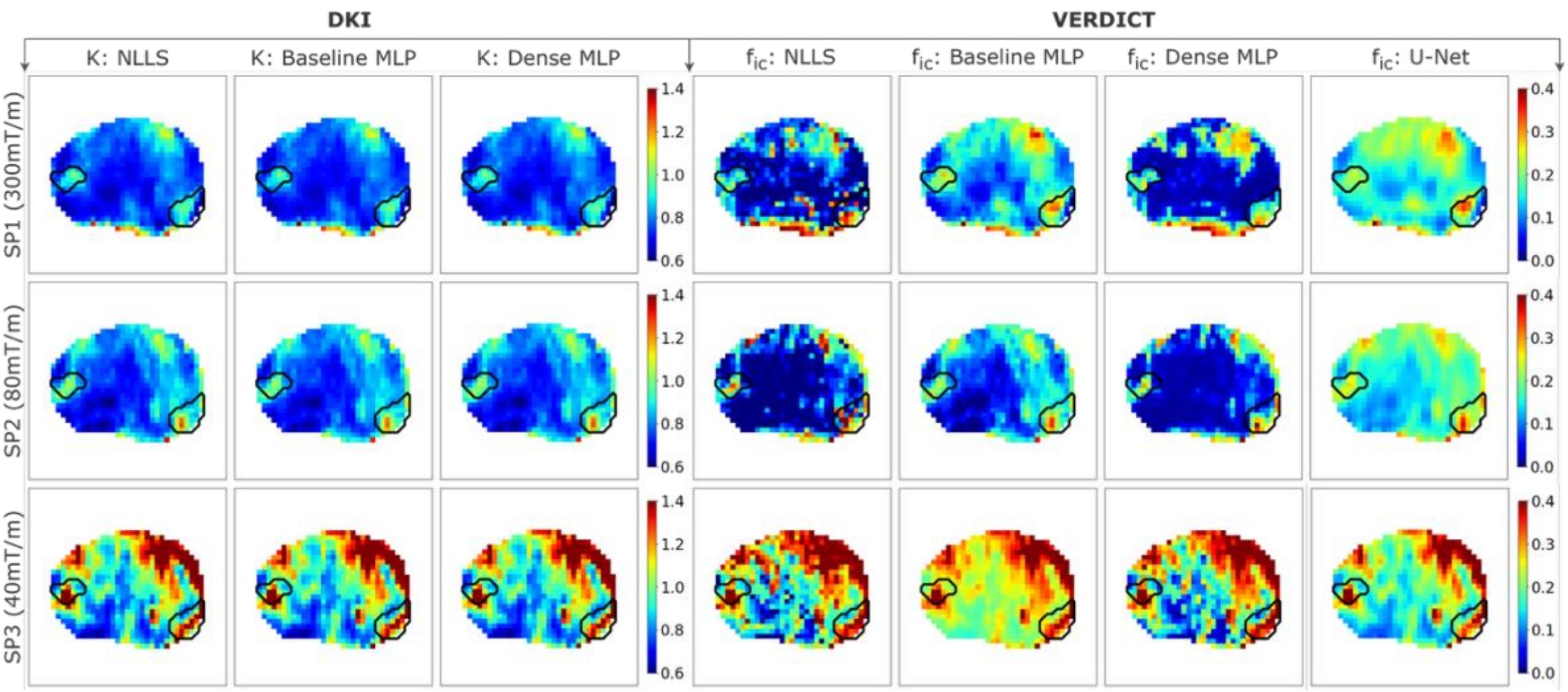

**Figure 6:** Comparison of $K$ and $f_{ic}$ parameter maps obtained using different fitting methods for both DKI and VERDICT fitting models across the three sub-protocols for Patient 3. There are two 3+4 Gleason grade tumours present, one at left posterior and lateral PZ, and the other at right posterolateral PZ. Both tumours are well visualized in maps obtained under SP1 across DKI methods, but conspicuity diminishes in the parameter maps corresponding to SP2 and SP3. VERDICT NLLS captures only the right lesion, whereas Baseline MLP, Dense MLP, and U-Net with strong gradients delineate both, with Dense MLP showing the best contrast. Maps derived from SP3 exhibit poor lesion visibility and artefacts, underscoring the limitations of low-gradient acquisitions (40 mT/m). For consistent visualization, $K$ values are constrained to the interval [0.6, 1.4], and $f_{ic}$ values are restricted to [0, 0.4], as indicated by the corresponding colorbars.

Figure 7 shows boxplots of the estimated $K$ and $f_{ic}$ distributions for Patient 3. Both DKI and VERDICT under SP1 and SP2 achieved strong separation between normal and tumour tissue ($p < 0.001$), whereas the distributions obtained under SP3 showed comparatively weaker differences ($p < 0.01$ for most methods, VERDICT NLLS $p < 0.05$). Overall, the parameter distributions obtained under SP1 were the narrowest and biophysically plausible.

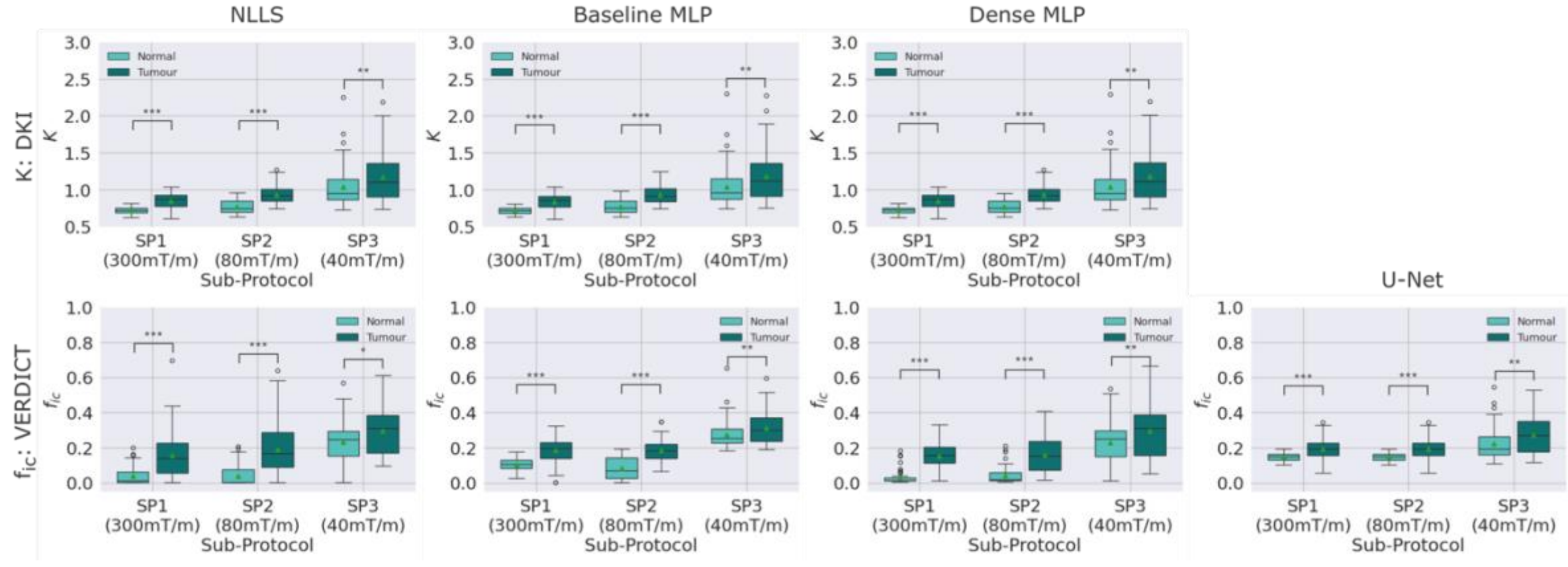

**Figure 7:** Boxplots illustrating the distributions of $K$ and $f_{ic}$ values across the three sub-protocols, compared across different fitting methods for both DKI and VERDICT fitting models for Patient 3. For DKI, data acquired under SP1 and SP2 show strongly significant differences ($p < 0.001$) between tumour and normal $K$, whereas data under SP3 show only moderate separation ($p < 0.01$). For VERDICT, SP1 and SP2 are associated with strong discrimination ($p < 0.001$) across all methods, while the distributions obtained under SP3 show comparatively weaker significance ($p < 0.05 - p < 0.01$). To ensure consistency in visualization, the $K$ values are constrained within the interval [0.5, 3.0], while the $f_{ic}$ values are restricted to [0, 1]. The following notation indicates statistical significance: $*** \rightarrow p < 0.001$; $** \rightarrow p < 0.01$; $* \rightarrow p < 0.05$; and *ns (non-significant)* → $p > 0.05$.

Table 6 summarizes the CoV, pooled SD of tumour $K$ and $f_{ic}$ values, and the CNR between normal and cancerous lesions for the ultra-strong gradient data (SP1). Clear differences were observed across models, with the VERDICT Dense MLP showing the most consistent behaviour. It achieved lowest inter-patient CoV and the lowest pooled voxel-wise SD within tumour ROIs, indicating most stable and reliable parameter estimates. Additionally, it produced the highest CNR median with the most compact IQR, reflecting strongest separation between tumour and contralateral normal tissue. Overall, these results suggest that the VERDICT Dense MLP provides the most robust and discriminative biomarker estimates among the evaluated models.

**Table 6:** SP1 (300 mT/m): Coefficient of Variation (CoV), pooled voxel-wise SD of tumour biomarker values, and CNR (median and IQR) between tumour and contralateral normal tissue across patient population for all diffusion MRI frameworks and fitting methods for the ultra-strong gradient data. The VERDICT Dense MLP achieves the strongest overall performance, exhibiting the lowest CoV, the lowest pooled SD, and the highest median CNR with the most compact IQR.

| Diffusion MRI Frameworks | Fitting Models | Coefficient of Variation (CoV) (%) | Standard Deviation: Tumour ROI Biomarker (Pooled) | Contrast-to-Noise Ratio (CNR) | |
|---|---|---|---|---|---|
| | | | | Median | IQR [Q1–Q3] |
| DKI ($K$) | NLLS | 48.10% | 7.87e-2 | 1.06 | 1.16 [0.94–2.21] |
| | Baseline MLP | 48.87% | 8.87e-2 | 1.10 | 1.33 [0.97–2.30] |
| | Dense MLP | 48.30% | 7.88e-2 | 1.08 | 1.33 [0.94–2.27] |
| VERDICT ($f_{ic}$) | NLLS | 43.76% | 1.29e-1 | 1.17 | 0.70 [1.00–1.70] |
| | Baseline MLP | 23.39% | 6.85e-2 | 1.30 | 1.56 [1.00–2.56] |
| | Dense MLP | **21.10%** | **6.48e-2** | **1.72** | **0.46 [1.40–1.86]** |
| | U-Net | 22.63% | 6.91e-2 | 1.02 | 1.13 [0.87–2.00] |

We also assessed the parameter maps and biomarker distribution boxplots for patients 1 and 4, with results demonstrated in Supplementary Information (Figures S1–S4). Additionally, we investigated whether any of the trained models, particularly VERDICT Dense MLP with access to ultra-strong gradient data, could meaningfully discriminate between the two Gleason grade groups (3+3 and 3+4). The boxplots for these Gleason groups are illustrated in Supporting Information (Figure S5).

# 4 Discussion

This work presents the first integration of ultra-strong gradient diffusion MRI with VERIDCT modelling for prostate cancer. Previous studies using ultra-strong gradients have primarily employed signal representations such as DKI [19] with conventional fitting approaches like NLLS, while VERDICT applications have so far been limited to clinical-strength systems [8, 11, 26, 31]. Our findings demonstrate that combining advanced acquisitions on high performance MRI systems with biophysical modelling enables new possibilities for non-invasive prostate tumour characterization.

We compared VERDICT against DKI and evaluated multiple fitting strategies, including traditional NLLS, a baseline self-supervised MLP [11], a denser MLP extending the baseline architecture, and a U-Net autoencoder. Our results show that VERDICT consistently outperformed DKI across all fitting methods, with the Dense MLP and NLLS achieving the lowest reconstruction error in different cases on the ultra-strong gradient data (SP1). VERDICT NLLS also yielded the lowest MSE and the highest AICc and BIC rankings across individual patients (PT 1, PT 2, PT 4, and PT 5; Tables 2–4, S1). This marginally better MSE for NLLS is expected, as it is explicitly optimized to minimize the residual and, in doing so, overfits to the local noise. This is reflected in the noisier parameter maps in Figures 4 and 6, and the higher inter-patient variability (CoV) and pooled SD reported in Table 6. Interestingly, U-Net underperformed relative to Dense MLP, showing higher MSE values and notably lower AICc and BIC rankings, likely due to its higher data requirements, as U-Net architectures typically require substantially larger training datasets to learn stable mappings. Importantly, ultra-strong gradients consistently reduced errors across sub-protocols, validating earlier findings that stronger diffusion gradients enhance sensitivity to restricted diffusion [32], and thereby improve characterization of the intracellular compartment.

In clinical practice, biomarkers reflecting tumour pathophysiology are vital for diagnosis and risk assessment, particularly in evaluating cancer grade and size. Consistent with our hypothesis, VERDICT Dense MLP with ultra-strong gradient data (SP1) demonstrated high tumour-to-normal conspicuity, supported by the highest CNR (with narrowest IQR) reported in Table 6. This enabled clear delineation even in patients with small tumours or poorly defined lesion boundaries where traditional methods were less effective. On visual inspection, the VERDICT Dense MLP produced $f_{ic}$ maps that were most anatomically similar to those from NLLS, but with notably reduced noise. U-Net achieved comparable tumour-normal conspicuity and surpassed NLLS and Baseline MLP, despite higher reconstruction losses. This highlights its ability to capture spatial microstructural features and suggests potential for further improvement with larger patient cohorts.

At subject level, the boxplot analysis of $K$ and $f_{ic}$ distributions showed that VERDICT Dense MLP under SP1 consistently delivered high level of statistical significance ($p < 0.001$) between tumour and normal tissues, particularly in cases with poor tumour conspicuity (PT 3 and PT 5). The U-Net outperformed both the NLLS and Baseline VERDICT models; however, differentiating the performance of VERDICT from DKI based on the boxplots alone remained challenging. Overall, the superior performance with SP1 underscores that ultra-strong gradients enhance sensitivity to cellular histology, enabling more reliable discrimination of tissue compartments. The enhanced lesion conspicuity and characterization achieved with VERDICT Dense MLP highlights its potential for integration with PI-RADS (Prostate Imaging Reporting and Data System) in the current clinical systems for scoring of indeterminate lesions, suggesting it could improve non-invasive tumour assessment and reduce the need for biopsies [7, 11, 33].

Despite the demonstrated strengths of VERDICT Dense MLP with SP1, Gleason grade discrimination revealed a limitation. Ultra-strong gradients did not consistently distinguish grade 3+3 from 3+4, with SP1 occasionally estimating higher mean $f_{ic}$ for 3+3, likely reflecting the small sample size rather than biological plausibility. Notably, SP3 (40 mT/m) showed the highest statistical significance in differentiating the grades among sub-protocols; however, this result is likely biased by the lower SNR at this gradient strength due to longer echo times [19] and should not be interpreted as a meaningful finding.

A limitation of this study was that the VERDICT dataset used constant diffusion parameters (Δ and δ) across all b-values, deviating from the variable diffusion time protocol in optimized VERDICT acquisitions. We applied the self-supervised learning framework to this constant-parameter acquisition. Additionally, the dataset exhibited voxel-wise gradient non-uniformity with spatially varying effective b-values. For neural network compatibility, we approximated these to the closest theoretical b-values for the entire volume, which did not account for spatial variations in diffusion weighting. Future studies should evaluate this framework on ultra-strong gradient datasets with optimized, variable diffusion time acquisitions and could explore methods to incorporate spatial gradient non-uniformity corrections into the network architecture, as recent work has demonstrated implicit neural representations can accommodate such spatial variations for diffusion microstructure modelling [34].

Another limitation of this study is the small cohort consisting of nine subjects that was not demographically matched due to limited sample size. Given that black men in the UK have nearly twice the risk of developing prostate cancer compared to white men [35, 36], limited or non-representative training data may hinder model generalizability and introduce potential ethnicity-related bias. The limited sample size also contributed to higher reconstruction loss for U-Net with VERDICT and non-convergence of DKI training. Although the training and validation curves were generally stable, minor inconsistencies such as lower validation losses than training losses likely arose from patient-specific variability introduced by the small cohort and the data-split strategy.

Expanding this work to a larger and more demographically diverse cohort in the future would be the most direct path to improving generalizability and robustness and to mitigate these limitations. Carefully incorporating data augmentation strategies could further enhance dataset diversity without compromising biophysical plausibility. With richer data, Vision Transformers (ViTs) [37] could be explored as a natural extension of the Dense MLP framework for more precise voxel-wise parameter estimation. Moreover, integrating relaxed-VERDICT (rVERDICT) [38] with ultra-strong gradient acquisitions may support improved Gleason-grade discrimination and could potentially contribute to future clinical applicability. Advanced deep learning-based uncertainty estimation methods like Variational Autoencoders (VAEs) [39] and Bayesian Neural Networks [40] could be incorporated to provide deeper insights into the reliability and robustness of ultra-strong gradient acquisition systems, particularly under noisy conditions.

# 5 Conclusion

This study demonstrates that integrating ultra-strong gradient diffusion MRI with self-supervised VERDICT models, particularly the Dense MLP architecture, yield accurate, robust, consistent, and physiologically meaningful estimation of prostate microstructural parameters. Across patients and sub-protocols relying on different gradient amplitudes (300, 80, and 40 mT/m), the VERDICT Dense MLP approach outperformed conventional DKI and NLLS fitting methods, offering superior in vivo tumour-normal tissue discrimination and improved lesion conspicuity with highest CNR. Although Gleason grade differentiation was limited by the small cohort, these findings highlight the potential of combining high-quality diffusion data with advanced deep learning to enhance non-invasive tumour characterization and reduce the need for unnecessary biopsies. The availability of commercial whole-body MRI systems equipped with ultra-strong gradients makes this demonstration timely, providing a pathway for translating the proposed methods from research settings into clinical use.

# 6 Acknowledgements

EP and SS are supported by the Department of Health's NIHR-funded Biomedical Research Centre at University College London Hospitals. MP is supported by a UKRI Future Leaders Fellowship (MR/T020296/2 and 1073) and the MRC Research Grant MR/W031566/1. CMWT is supported by a Vidi grant (21299) from the Dutch Research Council (NWO), and this research was funded in part by the Wellcome Trust [215944/Z/19/Z]. The UK National Facility for In Vivo MR Imaging of Human Tissue Microstructure in CUBRIC was funded by an EPSRC equipment grant (EP/M029778/1). We would like to explicitly acknowledge Malwina Molendowska for carrying out the data acquisition, and for sharing the dataset used in this work.

# 7 References

1. Bray F, Laversanne M, Sung H, et al. Global cancer statistics 2022: GLOBOCAN estimates of incidence and mortality worldwide for 36 cancers in 185 countries. CA: a cancer journal for clinicians. 2024;74(3):229-263.
2. Panagiotaki E, Walker-Samuel S, Siow B, et al. Noninvasive quantification of solid tumor microstructure using VERDICT MRI. Cancer research. 2014;74(7):1902-1912.
3. Afaq A, Andreou A, Koh DM. Diffusion-weighted magnetic resonance imaging for tumour response assessment: why, when and how? Cancer imaging. 2010;10(1A):S179.
4. Le Bihan D, Breton E, Lallemand D, Grenier P, Cabanis E, Laval-Jeantet M. MR imaging of intravoxel incoherent motions: application to diffusion and perfusion in neurologic disorders. Radiology. 1986;161(2):401-407.
5. Jensen JH, Helpern JA, Ramani A, Lu H, Kaczynski K. Diffusional kurtosis imaging: the quantification of non-gaussian water diffusion by means of magnetic resonance imaging. Magnetic Resonance in Medicine: An Official Journal of the International Society for Magnetic Resonance in Medicine. 2005;53(6):1432-1440.
6. Panagiotaki E, Schneider T, Siow B, Hall MG, Lythgoe MF, Alexander DC. Compartment models of the diffusion MR signal in brain white matter: a taxonomy and comparison. Neuroimage. 2012;59(3):2241-2254.
7. Johnston EW, Bonet-Carne E, Ferizi U, et al. VERDICT MRI for prostate cancer: intracellular volume fraction versus apparent diffusion coefficient. Radiology. 2019;291(2):391-397.
8. Bailey C, Bourne RM, Siow B, et al. VERDICT MRI validation in fresh and fixed prostate specimens using patient-specific moulds for histological and MR alignment. NMR in Biomedicine. 2019;32(5):e4073.
9. Bourne RM, Panagiotaki E, Bongers A, Sved P, Watson G, Alexander DC. Information theoretic ranking of four models of diffusion attenuation in fresh and fixed prostate tissue ex vivo. Magnetic resonance in medicine. 2014;72(5):1418-1426.
10. Vasylechko SD, Warfield SK, Afacan O, Kurugol S. Self-supervised IVIM DWI parameter estimation with a physics based forward model. Magnetic resonance in medicine. 2022;87(2):904-914.
11. Sen S, Singh S, Pye H, et al. ssVERDICT: Self-supervised VERDICT-MRI for enhanced prostate tumor characterization. Magnetic resonance in medicine. 2024;92(5):2181-2192.
12. Barbieri S, Gurney-Champion OJ, Klaassen R, Thoeny HC. Deep learning how to fit an intravoxel incoherent motion model to diffusion-weighted MRI. Magnetic resonance in medicine. 2020;83(1):312-321.

13. Gyori NG, Palombo M, Clark CA, Zhang H, Alexander DC. Training data distribution significantly impacts the estimation of tissue microstructure with machine learning. Magnetic resonance in medicine. 2022;87(2):932-947.
14. Epstein SC, Bray TJP, Hall-Craggs M, Zhang H. Choice of training label matters: how to best use deep learning for quantitative MRI parameter estimation. arXiv preprint arXiv:2205 05587. Published online 2022.
15. Huang HM. An unsupervised convolutional neural network method for estimation of intravoxel incoherent motion parameters. Physics in Medicine & Biology. 2022;67(21):215018.
16. Voorter PHM, Backes WH, Gurney-Champion OJ, et al. Improving microstructural integrity, interstitial fluid, and blood microcirculation images from multi-b-value diffusion MRI using physics-informed neural networks in cerebrovascular disease. Magnetic Resonance in Medicine. 2023;90(4):1657-1671.
17. Raissi M, Perdikaris P, Karniadakis GE. Physics-informed neural networks: A deep learning framework for solving forward and inverse problems involving nonlinear partial differential equations. Journal of Computational physics. 2019;378:686-707.
18. Karniadakis GE, Kevrekidis IG, Lu L, Perdikaris P, Wang S, Yang L. Physics-informed machine learning. Nature Reviews Physics. 2021;3(6):422-440.
19. Molendowska M, Palombo M, Foley KG, et al. Diffusion MRI in prostate cancer with ultra-strong whole-body gradients. NMR in Biomedicine. 2024;37(12):e5229.
20. Molendowska M, Mueller L, Fasano F, Jones DK, Tax CMW, Engel M. Giving the prostate the boost it needs: Spiral diffusion MRI using a high-performance whole-body gradient system for high b-values at short echo times. Magnetic Resonance in Medicine. 2025;93(3):1256-1272.
21. Cordero-Grande L, Christiaens D, Hutter J, Price AN, Hajnal JV. Complex diffusion-weighted image estimation via matrix recovery under general noise models. Neuroimage. 2019;200:391-404.
22. Lee HH, Novikov DS, Fieremans E. Removal of partial Fourier-induced Gibbs (RPG) ringing artifacts in MRI. Magnetic resonance in medicine. 2021;86(5):2733-2750.
23. Andersson JLR, Skare S, Ashburner J. How to correct susceptibility distortions in spin-echo echo-planar images: application to diffusion tensor imaging. Neuroimage. 2003;20(2):870-888.
24. Jovicich J, Czanner S, Greve D, et al. Reliability in multi-site structural MRI studies: effects of gradient non-linearity correction on phantom and human data. Neuroimage. 2006;30(2):436-443.
25. Bammer R, Markl M, Barnett A, et al. Analysis and generalized correction of the effect of spatial gradient field distortions in diffusion-weighted imaging. Magnetic Resonance in Medicine: An Official Journal of the International Society for Magnetic Resonance in Medicine. 2003;50(3):560-569.
26. Panagiotaki E, Chan RW, Dikaios N, et al. Microstructural characterization of normal and malignant human prostate tissue with vascular, extracellular, and restricted diffusion for cytometry in tumours magnetic resonance imaging. Investigative radiology. 2015;50(4):218-227.
27. Bonet-Carne E, Johnston E, Daducci A, et al. VERDICT-AMICO: Ultrafast fitting algorithm for non-invasive prostate microstructure characterization. NMR in Biomedicine. 2019;32(1):e4019.
28. Ronneberger O, Fischer P, Brox T. U-net: Convolutional networks for biomedical image segmentation. In: International Conference on Medical Image Computing and Computer-Assisted Intervention. Springer; 2015:234-241.
29. Grussu F, Schneider T, Zhang H, Alexander DC, Wheeler--Kingshott CAM. Neurite orientation dispersion and density imaging of the healthy cervical spinal cord in vivo. Neuroimage. 2015;111:590-601.
30. Tamura C, Shinmoto H, Soga S, et al. Diffusion kurtosis imaging study of prostate cancer: preliminary findings. Journal of magnetic resonance imaging. 2014;40(3):723-729.

31. Singh S, Rogers H, Kanber B, et al. Avoiding unnecessary biopsy after multiparametric prostate MRI with VERDICT analysis: the INNOVATE study. Radiology. 2022;305(3):623-630.
32. Genc S, Tax CMW, Raven EP, Chamberland M, Parker GD, Jones DK. Impact of b-value on estimates of apparent fibre density. Human brain mapping. 2020;41(10):2583-2595.
33. Turkbey B, Rosenkrantz AB, Haider MA, et al. Prostate imaging reporting and data system version 2.1: 2019 update of prostate imaging reporting and data system version 2. European urology. 2019;76(3):340-351.
34. Hendriks T, Arends G, Versteeg E, Vilanova A, Chamberland M, Tax CMW. Implicit neural representations for accurate estimation of the Standard Model of white matter. Communications Biology. Published online 2025.
35. Lloyd T, Hounsome L, Mehay A, Mee S, Verne J, Cooper A. Lifetime risk of being diagnosed with, or dying from, prostate cancer by major ethnic group in England 2008--2010. BMC medicine. 2015;13(1):171.
36. Bokhorst LP, Roobol MJ. Ethnicity and prostate cancer: the way to solve the screening problem? BMC medicine. 2015;13(1):179.
37. Zheng T, Yan G, Li H, et al. A microstructure estimation Transformer inspired by sparse representation for diffusion MRI. Medical Image Analysis. 2023;86:102788.
38. Palombo M, Valindria V, Singh S, et al. Joint estimation of relaxation and diffusion tissue parameters for prostate cancer with relaxation-VERDICT MRI. Scientific Reports. 2023;13(1):2999.
39. Xu M, Zhou Y, Goodwin-Allcock T, et al. MRI Parameter Mapping via Gaussian Mixture VAE: Breaking the Assumption of Independent Pixels. arXiv preprint arXiv:2411 10772. Published online 2024.
40. Jallais M, Palombo M. Introducing μGUIDE for quantitative imaging via generalized uncertainty-driven inference using deep learning. Elife. 2024;13:RP101069.

# 8 Supplementary Information

**Table S1:** SP1 (300 mT/m): Reconstruction MSE across diffusion MRI frameworks and fitting methods for individual patients under ultra-strong gradients. Patient 4 results are reported in Tables 2–4. VERDICT NLLS yielded the lowest losses for Patients 1, 2 and 5, while VERDICT Dense MLP preformed best for Patient 3. In all cases, these methods only marginally outperformed their next-best approaches.

| Diffusion MRI Frameworks | Fitting Models | Patient-wise MSE | | | |
|---|---|---|---|---|---|
| | | Patient 1 | Patient 2 | Patient 3 | Patient 5 |
| DKI | NLLS | 1.91e-3 | 2.29e-3 | 2.37e-3 | 8.23e-4 |
| | Baseline MLP | 1.96e-3 | 2.37e-3 | 2.41e-3 | 9.32e-4 |
| | Dense MLP | 1.92e-3 | 2.30e-3 | 2.37e-3 | 9.06e-4 |
| VERDICT | NLLS | **7.46e-4** | **9.33e-4** | 1.23e-3 | **3.76e-4** |
| | Baseline MLP | 9.13e-4 | 1.16e-3 | 1.38e-3 | 5.24e-4 |
| | Dense MLP | 7.49e-4 | 9.53e-4 | **1.20e-3** | 3.87e-4 |
| | U-Net | 2.49e-2 | 7.06e-3 | 3.64e-3 | 6.49e-4 |

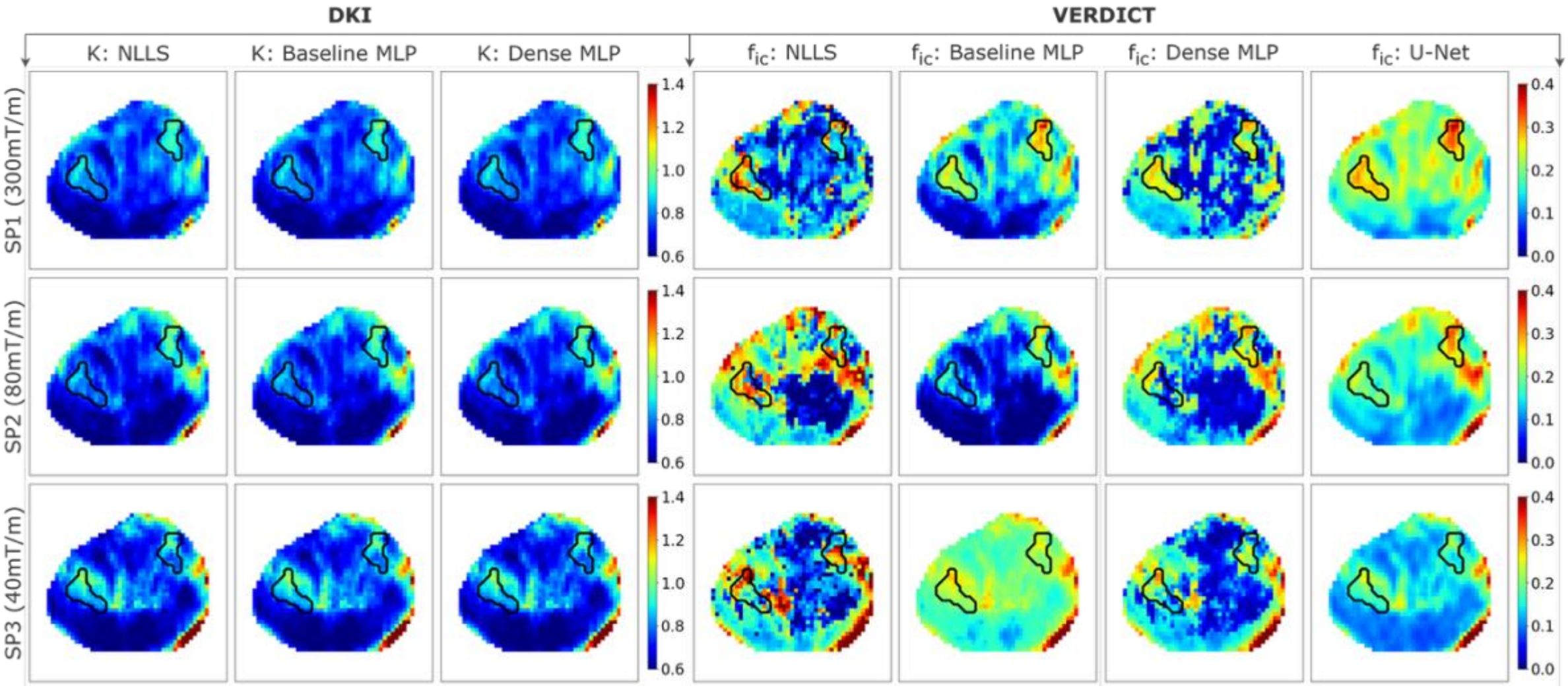

**Figure S1:** Comparison of $K$ and $f_{ic}$ parameter maps obtained using different fitting methods for both DKI and VERDICT fitting models across the three sub-protocols for Patient 1. Two 3+3 grade lesions are present: one at right posterolateral PZ and another at left anterior PZ, both at the apex. At 300 mT/m (SP1), VERDICT Baseline MLP, Dense MLP, and U-Net methods yield clear lesion delineation, whereas DKI and VERDICT NLLS remain less informative. The U-Net $f_{ic}$ map offers sharper delineation of the left anterior PZ tumour, whereas Dense MLP delineates right posterolateral PZ tumour with higher conspicuity. Lesion contrast diminishes in maps under SP2 and SP3. For consistent visualization, $K$ values are constrained to the interval [0.6, 1.4], and $f_{ic}$ values are restricted to [0, 0.4], as indicated by the corresponding colorbars.

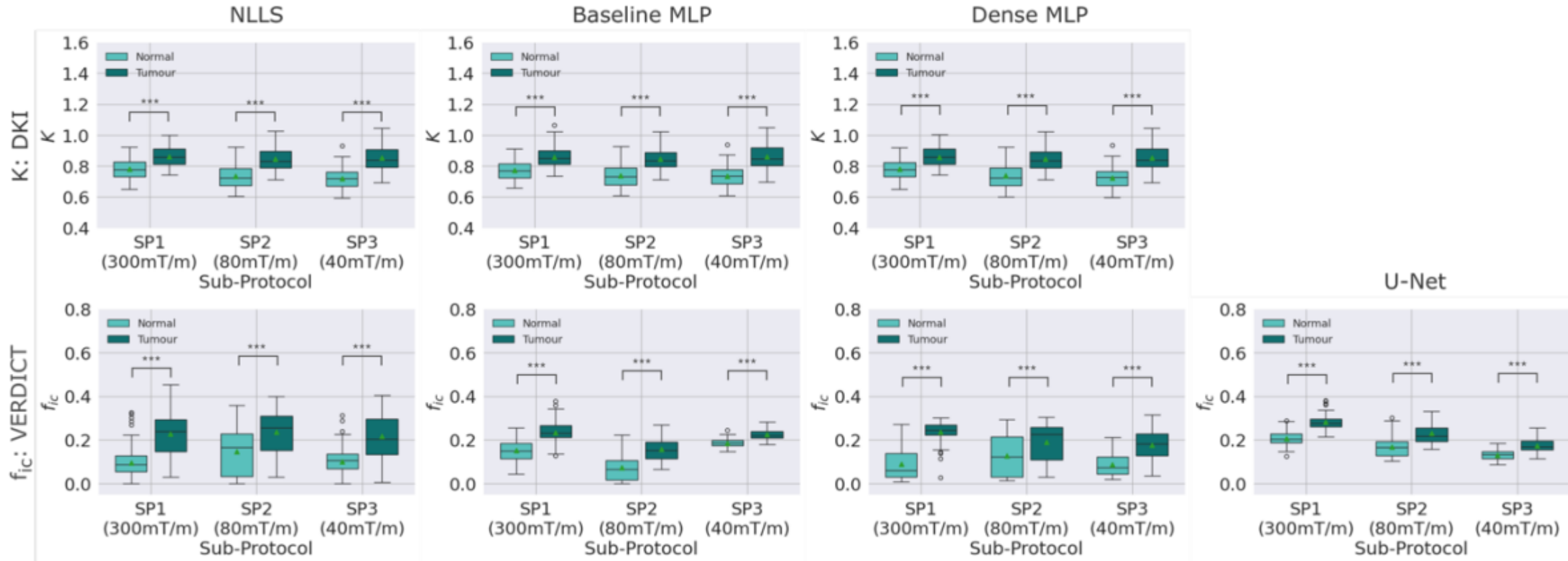

**Figure S2:** Boxplots illustrating the distributions of $K$ and $f_{ic}$ values across the three sub-protocols, compared across different fitting methods for both DKI and VERDICT fitting models for Patient 1. All DKI fitting methods show strong tumour-normal separation ($p < 0.001$). VERDICT similarly yields strong statistical difference ($p < 0.001$), with SP1 producing the narrowest distributions and clearest separation. To ensure consistency in visualization, the $K$ values are constrained within the interval [0.4, 1.6], while the $f_{ic}$ values are restricted to [0, 0.8]. The following notation indicates statistical significance: $*** \rightarrow p < 0.001$; $** \rightarrow p < 0.01$; $* \rightarrow p < 0.05$; and *ns (non-significant)* → $p > 0.05$.

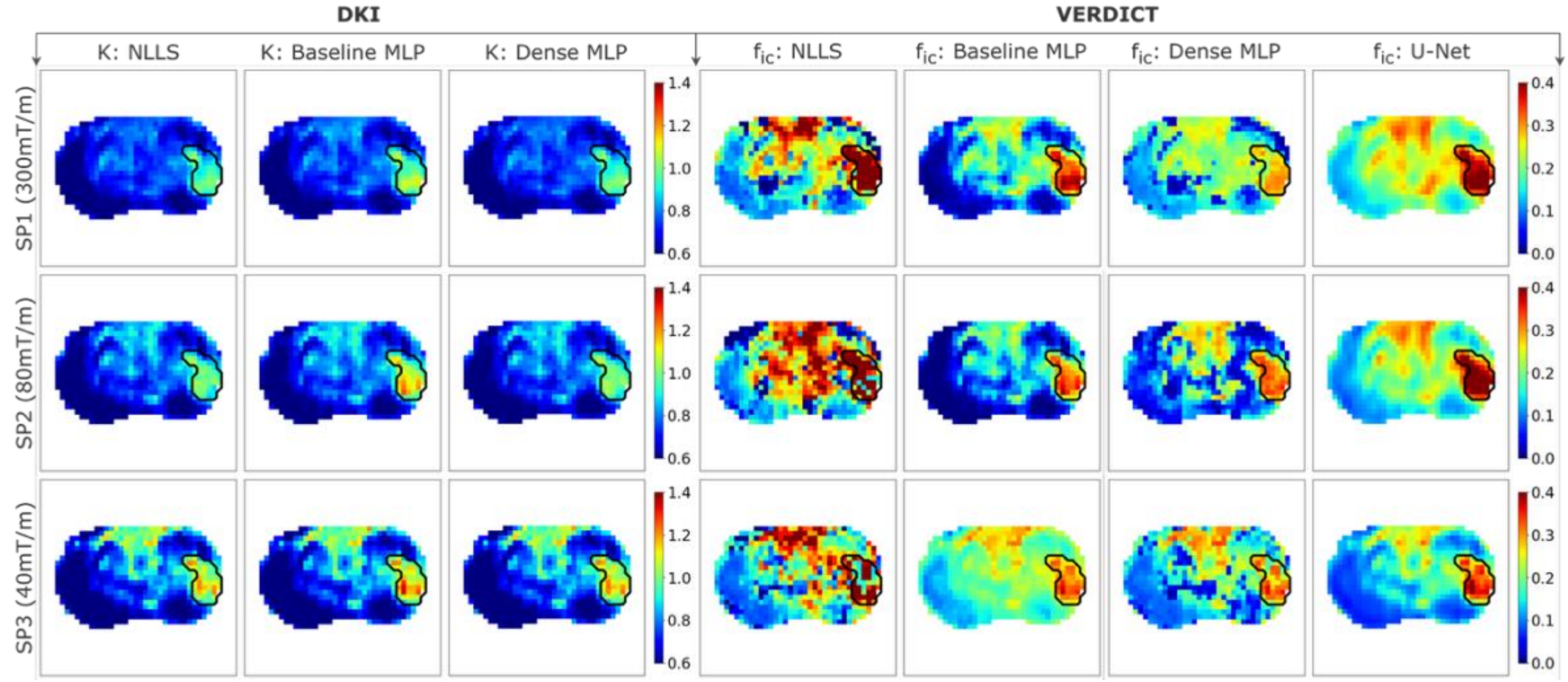

**Figure S3:** Comparison of $K$ and $f_{ic}$ parameter maps obtained using different fitting methods for both DKI and VERDICT fitting models across the three sub-protocols for Patient 4. A lesion of grade 3+4 is present at the left PZ at the base and mid-gland. All maps show clear tumour conspicuity. DKI maps appear similar across fitting methods, while VERDICT NLLS maps are noisy and U-Net produces the smoothest outputs. The VERDICT Baseline MLP underperforms due to $f_{ic}$ saturation near zero. The data acquired under SP1 and SP2 enable stronger lesion discrimination than those acquired under SP3, with maps from SP1 showing the best performance. For consistent visualization, $K$ values are constrained to the interval [0.6, 1.4], and $f_{ic}$ values are restricted to [0, 0.4], as indicated by the corresponding colorbars.

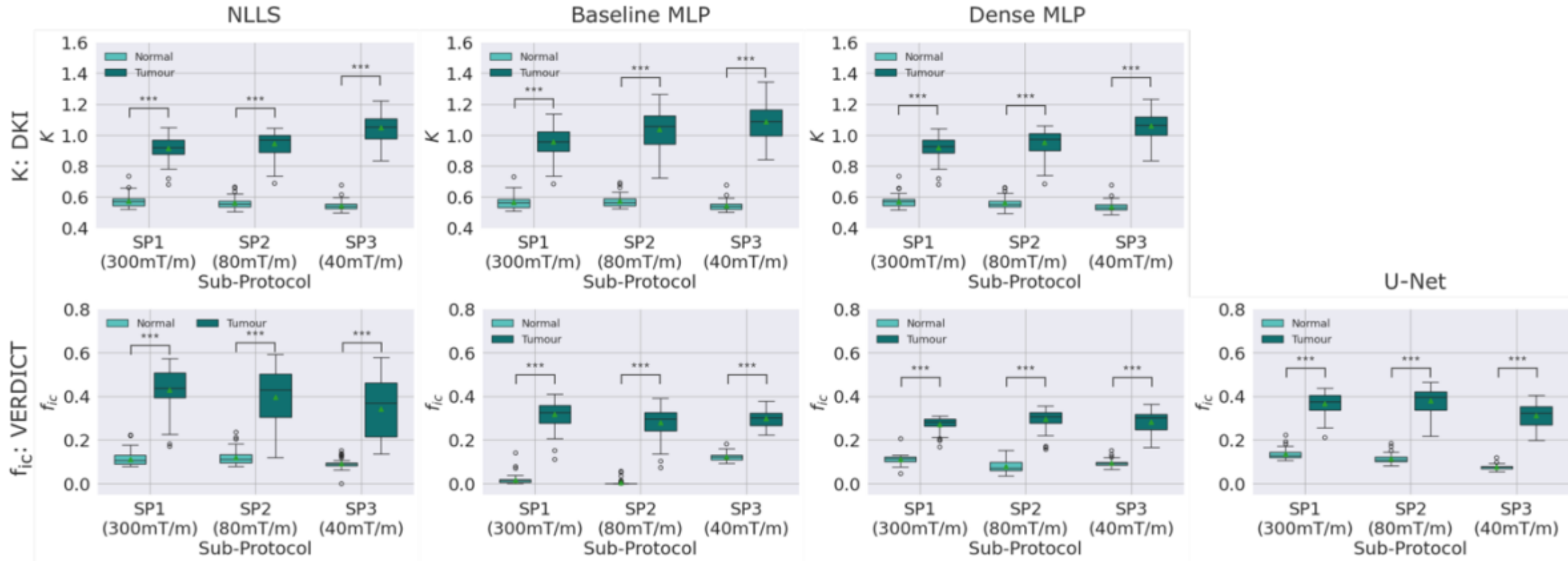

**Figure S4:** Boxplots illustrating the distributions of $K$ and $f_{ic}$ values across the three sub-protocols, compared across different fitting methods for both DKI and VERDICT fitting models for Patient 4. Both DKI and VERDICT show highly significant differences between tumour and normal regions ($p < 0.001$). VERDICT Baseline MLP produces implausible near-zero $f_{ic}$ estimates in normal ROIs, corrected by NLLS, Dense MLP, and U-Net. The data for SP1 (300 mT/m) yields the most concentrated distributions. To ensure consistency in visualization, the $K$ values are constrained within the interval [0.4, 1.6], while the $f_{ic}$ values are restricted to [0, 0.8]. The following notation indicates statistical significance: $*** \rightarrow\ p < 0.001$; $** \rightarrow\ p < 0.01$; $* \rightarrow\ p < 0.05$; and *ns (non-significant)* → $p > 0.05$.

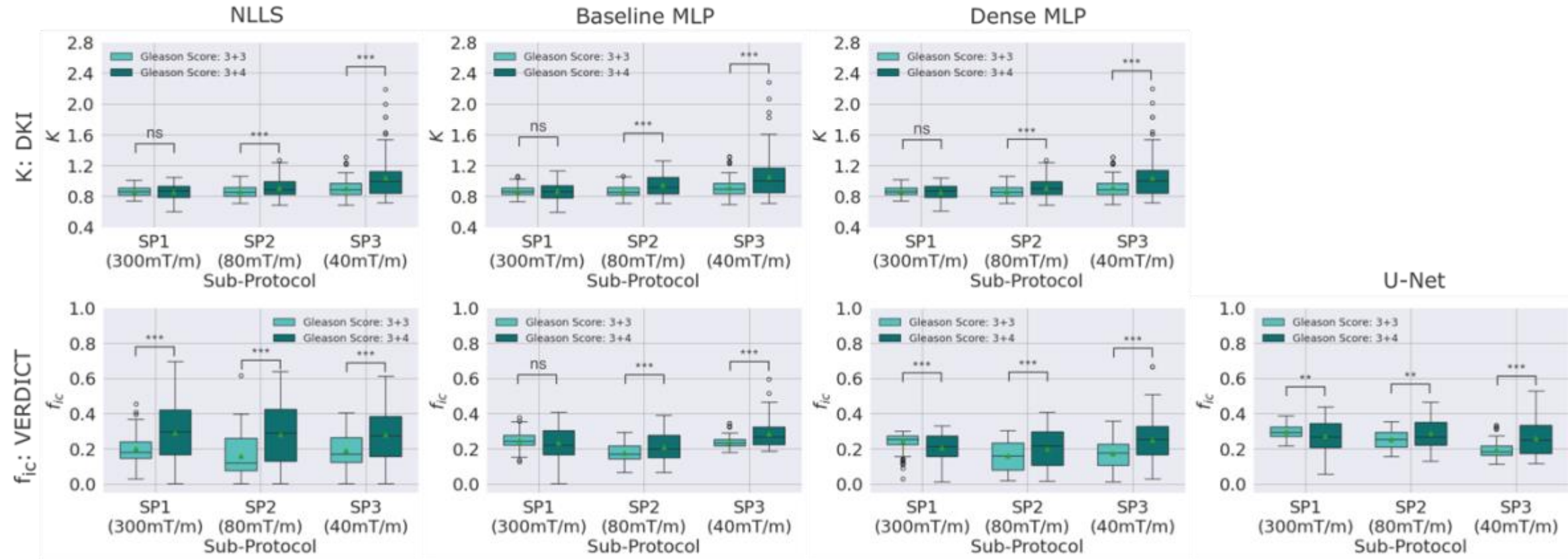

**Figure S5:** Boxplots illustrating the distributions of $K$ and $f_{ic}$ values across the three sub-protocols for Gleason scores 3+3 and 3+4, compared across different fitting methods for both DKI and VERDICT fitting models. Grade separation is not observed for DKI or the VERDICT Baseline MLP under SP1, whereas strong separation ($p < 0.001$) is observed under SP2 and SP3. Dense MLP achieved significant separation across all sub-protocols, though estimates under SP1 occasionally overestimated $f_{ic}$ in Gleason 3+3, contrary to histological expectations. Unexpectedly, the distributions obtained under SP3 provided the highest statistical Gleason-grade discrimination, which is likely a biased outcome driven by low SNR. To ensure consistency in visualization, the $K$ values are constrained within the interval [0.4, 2.8], while the $f_{ic}$ values are restricted to [0, 1]. The following notation indicates statistical significance: $*** \rightarrow\ p < 0.001$; $** \rightarrow\ p < 0.01$; $* \rightarrow\ p < 0.05$; and *ns (non-significant)* → $p > 0.05$.

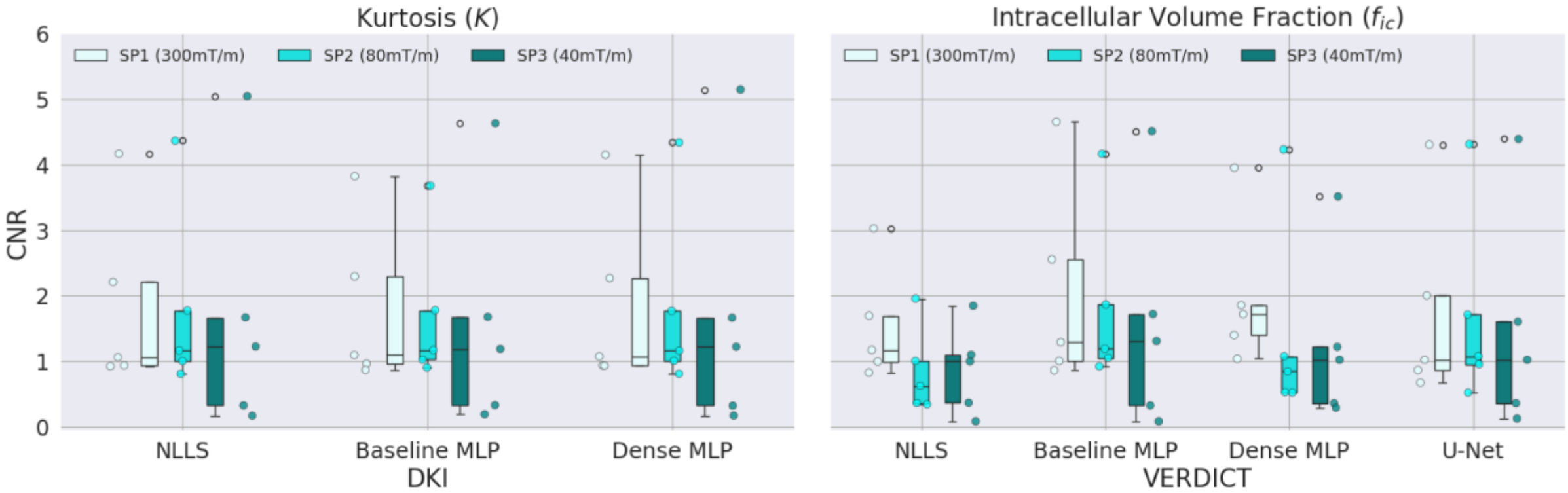

**Figure S6:** Boxplots showing CNR values across the three sub-protocols for different fitting methods for both DKI and VERDICT fitting models. Across all frameworks and fitting methods, VERDICT Dense MLP under SP1 (300 mT/m) achieves the highest median CNR and the narrowest inter-quartile range (IQR).

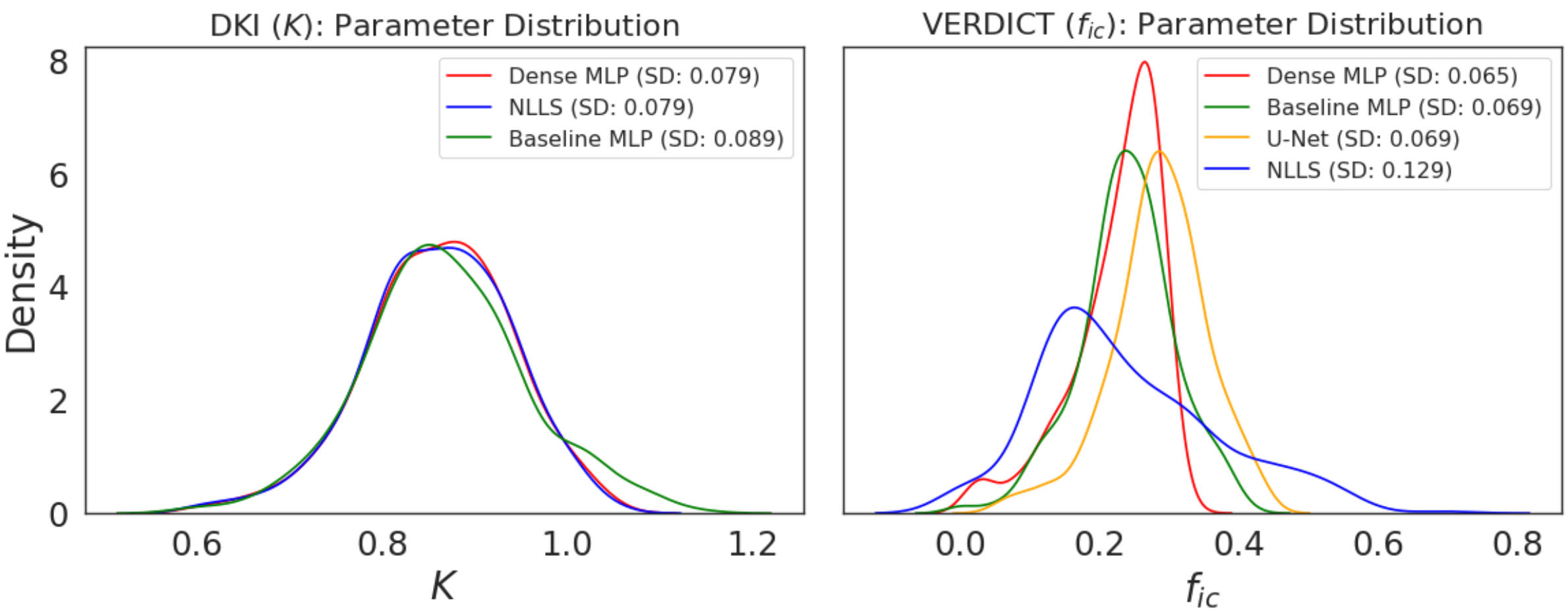

**Figure S7:** SP1 (300 mT/m): Density plots of $K$ and $f_{ic}$ parameters across diffusion MRI frameworks and fitting methods for all three sub-protocols, with the parameter values aggregated across patients. DKI methods produce highly similar distributions, whereas VERDICT distributions vary across fitting methods. VERDICT Dense MLP shows the lowest pooled standard deviation (SD), whereas VERDICT NLLS exhibits the highest pooled SD.